\documentclass[openacc]{rstransa}

\usepackage{bm}
\usepackage{color}
\usepackage{amsmath}

\newcommand{\beq}{\begin{eqnarray}}
\newcommand{\eeq}{\end{eqnarray}}

\begin{document}

\title{Multifaceted properties of Andreev bound states: Interplay of symmetry and topology}

\author{
T. Mizushima$^{1}$ and K. Machida$^{2}$}

\address{$^{1}$Department of Materials Engineering Science, Osaka University, Toyonaka, Osaka 560-8531, Japan \\
$^{2}$Department of Physics, Ritsumeikan University, Kusatsu 525-8577, Japan}

\subject{condensed matter physics, topological superconductors, quantum fluids}


\corres{Takeshi Mizushima\\
\email{mizushima@mp.es.osaka-u.ac.jp}}

\begin{abstract}
Andreev bound states ubiquitously emerge as a consequence of nontrivial topological structures of the order parameter of superfluids and superconductors and significantly contribute to thermodynamics and low-energy quantum transport phenomena. We here share the current status of our knowledge on their multifaceted properties as Majorana fermions and odd-frequency pairing. A unified concept behind Andreev bound states originates from a soliton state in the one-dimensional Dirac equation with mass domain wall, and the interplay of ABSs with symmetry and topology enriches their physical characteristics. We make an overview of Andreev bound states with a special focus on superfluid $^3$He. The quantum liquid confined to restricted geometries serves as a rich repository of noteworthy quantum phenomena, such as mass acquisition of Majorana fermions driven by spontaneous symmetry breaking, topological quantum criticality, Weyl superfluidity, and the anomalous magnetic response. The marriage of the superfluid $^3$He and nano-fabrication techniques will take one to a new horizon of topological quantum phenomena associated with Andreev bound states.

\end{abstract}


\begin{fmtext}
\section{Introduction}

Andreev bound states (ABSs) mirror the complexity of topological structure of order parameter manifold in superconductors (SCs) and superfluids (SFs)~\cite{kashiwayaRPP,tanakaJPSJ12,nagaiJPSJ08}. They ubiquitously emerge in topological defects such as vortices, surfaces, interfaces, and domain walls. The simplest example is a SC/normal/SC (SNS) junction. {The existence of the low-lying quasiparticles that are bound at the interface gives a fingerprint for the characteristic phase structure of two superconducting domains.} 
\end{fmtext}

\maketitle


\noindent
two superconducting domains. It is widely recognized that low-lying ABSs play fundamental roles in thermodynamics and quantum transport phenomena in low temperatures. In Pauli-limited SCs, furthermore, ABSs provide a key to capture an essence of the thermodynamic stability of the Fulde-Ferrell-Larkin-Ovchinnikov (FFLO) state~\cite{ff,lo,matsudaJPSJ07}, a self-organized periodic structure of superconducting gap. Indeed, surface and vortex ABSs have been directly observed in various kinds of SCs and SFs, including $^3$He~\cite{aokiPRL05, saitoh, wada, murakawaPRL09, murakawaJPSJ11,mizushimaJPCM15,mizushimaJPSJ15,okuda,volovik}, high-$T_{\rm c}$ cuprates~\cite{covington,kashiwaya,leseur}, and type-II SCs under a magnetic field~\cite{hess}. Recently, it has been recognized that the marriage of ABSs with symmetry and topology gives rise to a diversity of their characteristics and sheds light on new properties of ABSs, Majorana fermions and odd-frequency pair correlations. 

Majorana fermions are defined as {self-conjugate solutions of the Dirac equation} in quantum field theory~\cite{wilczek}. They are represented by the quantized fermionic field ${\bm \Psi}$ that satisfies the constraint of the self-charge-conjugation
\beq
\mathcal{C}{\bm \Psi}={\bm \Psi}, \hspace{5mm} \mathcal{C}^2=+1,
\label{eq:majorana}
\eeq
where $\mathcal{C}$ denotes the charge-conjugation operator or particle-hole operator in the context of condensed matter physics. Self-charge-conjugated Majorana fermions exist in topological and Weyl SCs/SFs as topologically protected gapless ABSs~\cite{tanakaJPSJ12,mizushimaJPCM15,mizushimaJPSJ15,qiRMP11}. It has recently been found that extra discrete symmetries adds a diversity of characteristics to such exotic quasiparticle, such as non-Abelian anyons subject to mirror reflection symmetry~\cite{ueno,sato14} and time-reversal symmetry~\cite{liuPRX14}, Majorana Ising spins in time-reversal-invariant SCs/SFs with magnetic point group symmetry~\cite{chungPRL09,nagatoJPSJ09,shindouPRB10,mizushimaPRL12,mizushimaPRB12,shiozakiPRB14}. The mirror-protected non-Abelian Majorana fermions are realized in $^3$He-A thin film, while the typical example of the latter is the SF $^3$He-B confined to a restricted geometry.

Another remarkable property of ABSs is odd-frequency Cooper pair correlations~\cite{tanakaJPSJ12,eschrigJLTP07,mizushimaJPCM15}. Anomalous charge and spin transport, electromagnetic responses, and proximity effects via ABSs have been clarified in light of odd-frequency Cooper pairing. The concept of odd-frequency pairing expands the classification of possible Cooper pairs. In accordance with the Fermi-Dirac statistics, the pairing symmetries of Cooper pair wave functions in a single-band SC can be classified into the fourfold way when the inversion symmetry is preserved~\cite{tanakaPRL07,tanakaPRL07-2,eschrigNP08}. Two of them are even-frequency spin-singlet even-parity (ESE) and even-frequency spin-triplet odd-parity (ETO) pairings, which do not change the sign of Cooper pair wave function by the exchange of times of paired fermions. There still remain two possibilities, odd-frequency spin-singlet odd-parity (OSO) and spin-triplet even-parity (OTE) pairs. Although there is no conclusive evidence on bulk odd-frequency superconductors~\cite{fominov,matsumoto,shigeta,hoshino}, {odd-frequency pair correlations ubiquitously appear in the topological defects of SCs and SFs as fluctuations of the condensate generated by pair-breaking and ABS formation~\cite{tanakaJPSJ12}.}


This article gives a review of recent progress on the interplay of symmetry and topology in ABSs with a special focus on SF $^3$He. In Sec.~\ref{sec:unified}, we start with a soliton state in the one-dimensional Dirac equation with a mass domain wall which offers a minimal model to capture an essence of gapless ABSs in SCs and SFs. The central part of this article is devoted to giving a unified description on the emergence of gapless ABSs on the basis of the soliton state and to clarify their connection with topology. We discuss the fundamental roles of ABSs on the spontaneous translational symmetry breaking in Pauli-limited SCs. In addition, the topological aspects of surface and vortex ABSs in Weyl SCs are uncovered, where the surface and vortex Fermi arc is topologically protected by the pairwise Weyl points. In Sec.~\ref{sec:symmetry}, after briefly introducing Majorana fermions, we show the current status of our knowledge on topological quantum phenomena woven by the intertwining of Majorana fermions with symmetry and topology. We here focus on the SF $^3$He as a rich repository of such topological phenomena. This includes mass acquisition of Majorana fermions driven by spontaneous symmetry breaking and topological quantum criticality. In addition, we clarify another property of ABSs {that the diagonal component of the Green's function constructed from ABSs is equivalent to the anomalous Green's function associated with odd frequency Cooper pair correlations. This clarifies that the formation of the ABS in SF $^3$He leads to the anomalous enhancement of surface spin susceptibilities~\cite{mizushimaPRB14}.} In Sec.~\ref{sec:conclusions} we give some prospects on searching exotic quasiparticles associated with the interplay of symmetry and topology in SFs and SCs.

Throughout this paper, we set $\hbar \!=\! k_{\rm B} \!=\! 1$ and the repeated Greek (Roman) indices imply the sum over $x, y, z$ (spins $\uparrow$ and $\downarrow$). The Pauli matrices in spin and particle-hole (Nambu) spaces are denoted by  $\sigma _{\mu}$ and $\tau _{\mu}$, respectively.



\section{A unified concept for Andreev bound states}
\label{sec:unified}

Let us start to give an overview on the topologically nontrivial structure of the one-dimensional Dirac equation with a spatially inhomogeneous mass $m(y)$, 
\beq
\left(
\begin{array}{cc}
m(y) & -i \partial _y \\
-i \partial _y & -m(y)
\end{array}
\right){\bm \varphi}_E(y) = E {\bm \varphi}_E(y).
\label{eq:dirac}
\eeq
Equation~\eqref{eq:dirac} enables to capture an essence of the Gross-Neveu model~\cite{gross} or Nambu-Jona-Lasinio model in $1+1$ dimensions~\cite{nambuPR1961} which is the central part of a renormalizable quantum field theory for interacting fermions with $N$ flavors. Although the model holds the discrete or continuous chiral symmetry, spontaneous chiral symmetry breaking in the vacuum gives rise to dynamical mass acquisition of fermions, $m\neq 0$, which is characterized by the auxiliary fields composed of pairwise fermions. For spatially uniform $m$, the fermion has a finite energy gap $\min |E|=|m|$.

Jackiw and Rebbi \cite{jackiw} clarified that for the boundary condition $m(y\rightarrow \pm \infty) ={\rm const.}$, the eigenfunction of the zero energy state is obtained by integrating Eq.~(\ref{eq:dirac}) with $E = 0$ as
\beq
{\bm \varphi}_{E=0}(y) = N \exp\left( 
- \int^{y}_0 m (y^{\prime})dy^{\prime}
\right)\left( 
\begin{array}{c}
1 \\ i
\end{array}
\right),
\label{eq:diracwf}
\eeq
where $N$ is a normalization constant. {We choose the sign of the exponent in Eq.~\eqref{eq:diracwf} so that the normalization requires ${\rm Re}m>0$ for the $y>0$ regin.} Since the mass $m(y)$ approaches a uniform value in the limit of $y\rightarrow \pm \infty$, the zero energy solution is normalizable only when the mass term satisfies the condition
\beq
\arg m(+\infty) - 
\arg m(-\infty) 
= (2n+1) \pi,
\label{eq:piphase}
\eeq
where $n\!\in\! \mathbb{Z}$. This indicates that, in addition to the continuum state with $|E|>|m(\pm\infty)|$, there exits at least one zero energy solution when the mass term $m (y)$ changes its sign at $y\rightarrow\pm\infty$, whose wave function is tightly bound to the mass domain wall. The stability of the zero energy state is independent of the detailed structure of $m(y)$. 

Recently, general solutions of Eq.~(\ref{eq:dirac}) have been found by using a technique of the Ablowitz-Kaup-Newell-Segur hierarchy well-known in integrable systems~\cite{akns}. These include a single kink state of the mass $m(y)$~\cite{hu,bar-sagi,shei}, multiple kinks (kink-anti-kink and kink-polaron states)~\cite{campbell,okuno,feinberg2003,feinberg2004}, and complex kinks and their crystalline states~\cite{basar1,basar2,basar3,takahashi}. These nontrivial kink and crystalline structures are accompanied by low-lying solitonic fermions and band structures formed by soliton lattice.

In superconducting states, electronic states are determined by the Bogoliubov-de Gennes (BdG) Hamiltonian, which is given in the basis of the Nambu spinor ${\bm \psi}=(\psi _{\uparrow},\psi _{\downarrow},\psi^{\dag}_{\uparrow},\psi^{\dag}_{\downarrow})^{\rm t}$ as
\beq
\mathcal{H}({\bm k},{\bm r}) = \left( 
\begin{array}{cc}
\varepsilon ({\bm r}) & \Delta ({\bm k},{\bm r}) \\
-\Delta^{\ast}(-{\bm k},{\bm r}) & -\varepsilon^{\rm t}({\bm r})
\end{array}
\right).
\label{eq:bdgH}
\eeq
The matrix $\varepsilon({\bm r})$ describes the single-particle Hamiltonian in the normal state and the $2 \times 2$ gap function $\Delta({\bm k},{\bm r})$ satisfies $\Delta({\bm k},{\bm r})=-\Delta(-{\bm k},{\bm r})^{\rm t}$. The BdG Hamiltonian is diagonalized in terms of the energy eigenstates as 
\beq
{\bm \psi}({\bm r})=\sum _{E>0}\left[
{\bm \varphi}_E({\bm r}){ \eta}_E+\mathcal{C}{\bm \varphi}_E({\bm r})\eta^{\dag}_E
\right], 
\label{eq:psi}
\eeq
where $\eta _E$ and $\eta^{\dag}_E$ denote the annihilation and creation operators of quasiparticles with $E$, respectively. The particle-hole operator is given as $\mathcal{C}=\tau _x K$ with the complex conjugation operator $K$. The BdG Hamiltonian maintains the particle-hole symmetry, $\mathcal{C}\mathcal{H}({\bm k},{\bm r})\mathcal{C}^{-1}=-\mathcal{H}(-{\bm k},{\bm r})$. 
The energy eigenstates in superconducting states are obtained by solving the BdG equation, $\int d{\bm r}_{2}\mathcal{H}({\bm r}_1,{\bm r}_2){\bm \varphi}_E({\bm r}_2)=E{\bm \varphi}_E({\bm r}_1)$. By employing the Andreev approximation where $\Delta$ is much smaller than the Fermi energy $E_{\rm F}$, the normal state dispersion is reduced to $\varepsilon({\bm r}) \approx -i {\bm v}_{\rm F}(\hat{\bm k})\cdot{\bm \nabla}$  (${\bm v}_{\rm F}$ denotes the Fermi velocity) and ${\bm \varphi}$ is decomposed to the slowly varying part $\tilde{\varphi}$ and the rapid oscillation part with the Fermi wave length $k^{-1}_{\rm F}$. The slowly varying function is governed by the Andreev equation
\beq
\left[
-i {\bm v}_{{\rm F}}(\hat{\bm k}) \cdot {\bm \nabla} {\tau}_z  + \underline{\Delta} (\hat{\bm k},{\bm r})
\right]\tilde{\bm \varphi}_{E}({\bm r}) = E \tilde{\bm \varphi}_E({\bm r}),
\label{eq:a}
\eeq
{where}
\beq
{\underline{\Delta} (\hat{\bm k},{\bm r})=\begin{pmatrix}
0 & \Delta (\hat{\bm k},{\bm r}) \\ \Delta (\hat{\bm k},{\bm r})^{\dag} & 0
\end{pmatrix}}.
\eeq
We will see below that the Andreev equation for unconventional SCs and SFs can be mapped onto the one-dimensional Dirac equation \eqref{eq:dirac} with an effective mass $m$, which gives a unified description for low-lying quasiparticle states bound to defects, including surface and vortex ABSs. We also emphasize the non-trivial topological aspect of such bound states.

\subsection{Fulde-Ferrell-Larkin-Ovchinnikov states}
\label{sec:fflo}

In the context of condensed matter physics, Eq.~\eqref{eq:dirac} serves as a unified description for low-lying electronic states in various ordered systems with self-organized periodic structure. This includes the one-dimensional Peierls system~\cite{brazovskii,mertsching,horovitz}, spin density waves~\cite{machidaPRB1984-2}, the spin-Peierls system~\cite{fujitaJPSJ1984}, the stripes in high-$T_{\rm c}$ cuprates~\cite{machida1989}, superconducting junction systems~\cite{kashiwayaRPP}, and Fulde-Ferrell-Larkin-Ovchinnikov (FFLO) states~\cite{machidaPRB1984,yoshii}. We here discuss the fundamental roles of ABSs on the thermodynamic stability of the FFLO state in a spin-singlet Pauli-limited SC.

For one-dimensional spin-singlet $s$-wave SCs, $\Delta ({\bm k},{\bm r})=\Delta (y)i\sigma _y$, the Andreev equation \eqref{eq:a} is reduced to the Dirac equation \eqref{eq:dirac} with an appropriate unitary transformation of the spin Pauli matrices $\sigma _{\mu}$, where fermions acquire an effective mass associated with the superconducting gap. The condition \eqref{eq:piphase} indicates that a zero energy solitonic state exists when two asymptotic superconducting gaps have the $\pi$-phase shift,
\beq
{{\rm arg}\Delta(+\infty)-\arg\Delta(-\infty)=(2n+1)\pi.}
\label{eq:pi}
\eeq
In other words, ABSs emerge at the interface of a SNS junction, whose energy is characterized by $E\propto \cos(\varphi/2)$ with an arbitrary phase shift $\varphi$~\cite{kashiwayaRPP}. The zero energy solitonic state is therefore regarded as a special kind of ABSs in SNS junctions ($\varphi = \pi$).

Since the superconducting gap is constructed from pairwise quasiparticles, the BdG or Andreev equation is supplemented by the gap equation, which gives a closed set for describing self-consistent quasiparticle structures in spatially inhomogeneous SCs/SFs. The simplest form that satisfies the condition \eqref{eq:pi} is a single kink solution, 
\beq
\Delta (y)=\Delta _0 \tanh(y/\xi),
\label{eq:kink}
\eeq
where $\xi=v_{\rm F}/\Delta _0$ is the superconducting coherence length. The generic solution of the self-consistent equations is obtained in terms of Jacobi elliptic functions~\cite{machidaPRB1984}, which contains Eq.~\eqref{eq:kink} as an exact solution. 

\begin{figure}[!t]
\centering\includegraphics[width=5.2in]{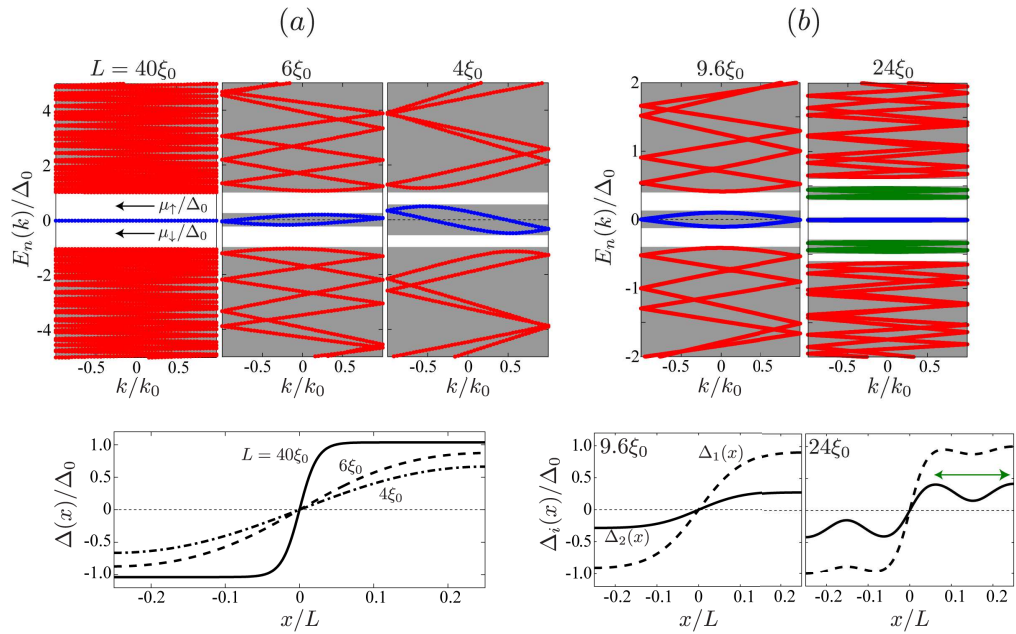}
\caption{(a) Quasiparticle spectra in the reduced Brillouin zone (top) and self-consistent superconducting gap $\Delta (x)$ (bottom) in a single-band FFLO state for various modulation periods $L=4$, $6$, and $40\xi _0$, where $\mu _{\uparrow,\downarrow}$ denote the Fermi surface of spin $\uparrow,\downarrow$ electrons {and we set $k_0\equiv \pi /L$.} As $L/\xi _0$ decreases, {the continuum states (red) essentially remain unchanged,} while the zero energy ABSs {(blue)} become dispersive. (b) Quasiparticle spectra for the minority band (top) and $\Delta _{1,2}(x)$ (bottom) in the different subphases of two-band FFLO states. The gap $|\Delta _1|$ in the majority ``1'' band is larger than $|\Delta _2|$, and we set the ratio of the Fermi velocity as $v_{{\rm F},1}/v_{{\rm F},2}>1$. The spatial modulation of $\Delta _{1,2}(x)$ in $L=9.6\xi_0$ can be characterized by a single modulation vector $Q$. A multiple-Q modulation structure emerges in $L=24\xi_0$, {where $Q_1$ is defined by $L$ and $Q_2$ characterizes the shorter modulation shown in the arrow. The resulting quasiparticle spectrum has the characteristic ABSs (green).}}
\label{fig:fflo}
\end{figure}

The single kink solution describes the domain wall of two degenerate vacua $\pm \Delta _0$ and the condition~\eqref{eq:pi} ensures the existence of the zero energy state localized at the domain wall. The kink-shaped superconducting gap also generates phase shifts of continuum states. Owing to the presence of a soliton and phase shifts, the creation of the single kink costs the energy $E[\Delta(y)]-E[\Delta _0]=\frac{2}{\pi}\Delta _0$~\cite{takayama} and thus the kink state can not be thermodynamically stable at zero fields. The energy cost is, however, compensated by the Pauli paramagnetic energy, $\mu _{\rm B}H$, since the splitting of Fermi surfaces generated by a magnetic field accommodate an excess spin into the solitonic $E=0$ state. The kink structure is therefore stabilized in the high field regime, $H>H_{\rm LO}\equiv\frac{2}{\pi \mu _{\rm B}}\Delta _0$, which is lower than the Pauli limiting field $H_{\rm P}\equiv \Delta_0/\sqrt{2}\mu_{\rm B}$. The field, $H_{\rm LO}$, is known as the lower critical field to stabilize the FFLO state in Pauli-limited SCs. As the magnetic field increases, the superconducting gap $\Delta (y)$ forms self-organized periodic structure of multiple kink-antikink pairs and the quasiparticle spectrum possesses the zero energy flat band which accommodate excess spins~\cite{machidaPRB1984,mizushimaPRL05-1}, as displayed in Fig.~\ref{fig:fflo}(a). In the high field regime, the order parameter profile is transformed to a sinusoidal shape with a single modulation vector $Q\sim \mu _{\rm B}H/v_{\rm F}$. In this regime, the zero energy solitonic states bound to each nodal point of $\Delta (y)\sim\Delta _0 \sin(Qy)$ interfere with their neighbors and the zero energy flat band turns to the dispersive band structure due to the formation of a soliton lattice within the energy scale $\sim {\rm e}^{-L/2\xi}$, where $L=2\pi /Q$ is a period of the FFLO modulation, $\Delta (x+L)=\Delta (x)$. The formation of a self-organized periodic structure in the FFLO state is a direct consequence of the synergistic effect between spin paramagnetism and superconductivity with the spontaneous breaking of the translational symmetry, and the solitonic state provides a key concept to capture the essence. 

Although the ABSs in FFLO SCs are not topologically protected, they possess the characteristics of odd-frequency spin-singlet odd-parity Cooper pair correlations due to the spontaneous breaking of the translational symmetry~\cite{yokoyamaJPSJ10,higashitaniPRB14}. {It has been discussed that ABSs play a key role on the observability of a signature of the FFLO modulation through the local density of states and modulated magnetization density~\cite{wang,anton,ichiokaPRB07,suzukiJPSJ11}. In unconventional SCs, such as a two-dimensional $d$-wave SC, the interplay of ABSs with the nodal pairing leads to the reorientation of the $Q$-vector from the nodal direction to anti-nodal direction in a high field~\cite{anton}.}

For Pauli-limited two-band SCs, as a result of the competing effect between two bands, the FFLO phase is divided by successive first order transitions into an infinite family of FFLO subphases with rational modulation vectors~\cite{mizushimaJPSJ14,takahashiPRB14}. When two electron bands have different superconducting gaps $|\Delta _1| \neq |\Delta _2|$ and Fermi velocities $v_{{\rm F},1}\neq v_{{\rm F},2}$, each band has its own favorite FFLO modulation vector $Q_1\neq Q_2$. In Fig.~\ref{fig:fflo}(b), we display the different FFLO subphases in two-band SCs, where $|\Delta _1|>|\Delta _2|$ and $v_{{\rm F},1}>v_{{\rm F},2}$. The set of parameters indicates that in the vicinity of the FFLO lower critical field, the FFLO modulation period in the minority ``2'' band, $Q^{-1}_2$, is comparable to the coherence length $Q^{-1}_2\!\sim \! \xi _0$ and $Q^{-1}_1 \!\gg\! \xi _0$. These two modulation periods start to compete with each other, when the Cooper pair tunneling between two bands is switched on. It is recently found in Ref.~\cite{mizushimaJPSJ14} that as a consequence of the competition, the overall modulation of the superconducting gap is characterized by a rational modulation vector of $Q_2$, such as $Q=Q_2/(2n+1)$ ($n\in\mathbb{Z}$). The FFLO modulation is then expanded in terms of the higher harmonics $(2m+1)Q$ as $\Delta _{i}(x) = \sum _m\Delta^{(m)}_ie^{i(2m+1)Qx} = \Delta^{(1)}_i e^{iQ_2x/(2n+1)} + \cdots + \Delta^{(2n+1)}_i e^{iQ_2x} + \cdots$, which contains the favorable modulation vector $Q_2$ in addition to the long wavelength $Q_2/(2n+1)\sim Q_1$. Hence, the resultant phase diagram realized in two-band Pauli-limited SCs may have FFLO subphases characterized by the series of the rational number $Q_2/(2n+1)$. Figure \ref{fig:fflo}(b) with $L=24\xi _0$ shows the typical subphase with $n=3$. The original band in the FFLO subphase with $Q=Q_2/(2n+1)$ is folded back into a small reduced Brillouin zone by the mixed component with the larger modulation vector $(2n+1)Q=Q_2$. Then, the folding back generates new electron bands below the threshold energy of the continuum $|E|<\Delta _0$ as displayed in the quasiparticle spectrum of Fig.~\ref{fig:fflo}(b) (green). The multiple-$Q$ modulation structure in the FFLO subphases can be clearly reflected by a devil's staircase structure in the field dependences of the modulation vector and paramagnetic moment~\cite{mizushimaJPSJ14}. These features inherent to multiband SCs are distinct from those in single-band systems where the FFLO state emerges via the second-order transition and no subphases exist.

\subsection{Surface Andreev bound states in Weyl superconductors}

Let us now apply Eq.~\eqref{eq:piphase} to surface ABSs emergent in unconventional SCs. As a pedagogical model, we here consider a chiral $\ell$-wave pairing state, where $\ell \!=\! 1$, $2$, and $3$ correspond to chiral $p$-, $d-$, and $f$-wave, respectively. We also do not take account of spin degrees of freedom and spatial uniformity of $\Delta$ is assumed. The superconducting gap is then given by
\beq
\Delta ({\bm k},{\bm r}) = \Delta _0 (\hat{k}_x + i\hat{k}_y)^{\ell},
\label{eq:deltaell}
\eeq
where $\hat{\bm k}\equiv {\bm k}/|{\bm k}| \approx {\bm k}/k_{\rm F}$. Owing to the spontaneous breaking of the time-reversal symmetry, the chiral state ($\ell \ge 1$) in three dimensions is always accompanied by Fermi points at which the bulk quasiparticle excitation is gapless. The particle-hole symmetry of the BdG Hamiltonian guarantees the presence of pairwise Fermi points at ${\bm k}={\bm k}_0$ and ${\bm k}=-{\bm k}_0$. 

\begin{figure}[!t]
\centering\includegraphics[width=5.2in]{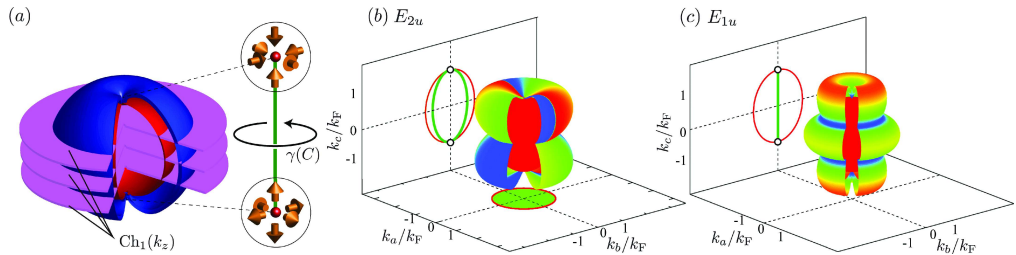}
\caption{(a) Topological structure of chiral $p$-wave pairing: pairwise Weyl points, Berry phase $\gamma (C)$ around the Dirac string, and the first Chern number ${\rm Ch}_1$ on each sliced momentum plane. The thick arrows denote the unit vector $\hat{\bm m}({\bm k})$. Superconducting gap and surface Fermi arcs in the $E_{2u}$ state (b) and $E_{1u}$ state (c) as possible candidates for the B-phase of UPt$_3$.}
\label{fig:weyl}
\end{figure}

Volovik~\cite{volovik} found that in chiral pairing states, the Fermi points are protected by the nontrivial first Chern number defined on a two-dimensional plane embracing the Fermi points. The pairwise Fermi points, which are called the Weyl points, can not be removed by any perturbations. The effective Hamiltonian around the Weyl points is mapped onto the Weyl Hamiltonian, $\mathcal{H}=\hat{\bm m}({\bm k})\cdot{\bm \sigma}$, and low-lying quasiparticles bound to the Fermi points behave as Weyl fermions. Weyl fermions are massless fermions expressed by a two-dimensional spinor with a well-defined notion of the left- or right-handed coordinate, i.e., chirality. Owing to the particle-hole symmetry, the Weyl fermions risiding on the pairwise Weyl points have opposite chirality. In the context of the Berry phase, the Weyl points are regarded as ``magnetic monopoles'' in momentum space and the pairwise Weyl points are connected by a Dirac string which is accompanied by the Berry phase change $\gamma(C)=2\ell\pi$ along a closed loop ``$C$'' embracing the string (see Fig.~\ref{fig:weyl}(a)). The first Chern number ${\rm Ch}_1$ defined in each sliced two-dimensional momentum plane is found to be ${\rm Ch}_1(k_z)=\ell$ for $|k_z| \!<\! k_{\rm F}$, and otherwise it is trivial. Hence, three-dimensional SCs concomitant with Weyl points can be regarded as a layered structure of particle-hole symmetric quantum Hall states. The bulk-edge correspondence guarantees the existence of the zero energy states in each momentum plane. They form the zero energy flat band structure along the ${\bm k}$ direction connecting two Weyl points. Hence, the Fermi points of Weyl SCs correspond to the end point of the zero energy flat band, leading to the ``Fermi arc'' protected by pairwise Weyl points~\cite{silaev:2012,mizushimaJPSJ15}. The presence of the topologically protected Fermi arc is responsible for a pronounced zero-bias conductance peak in tunneling spectroscopy and may significantly affect quantum transport phenomena. 

When the chemical potential is negative, the quasiparticle spectrum in the chiral state \eqref{eq:deltaell} is fully gapped and surface Fermi arcs disappear. In this situation, ${\rm Ch}_1$ that protects the pairwise Fermi points becomes trivial and thus the Weyl points vanish. There exists the topological phase transition at $\mu=0$, where the bulk excitation becomes gapless~\cite{read,mizushimaPRL08}.

The chiral pairing state in Eq.~\eqref{eq:deltaell} can be a prototype of Weyl SCs, superconducting analogue to Weyl semimetals~\cite{balentsPRB12,sauPRB12,dasPRB13,yangPRL14,silaevJETP14}. The concrete example is the superfluid $^3$He-A confined to a thin film. The superfluid is known as the Anderson-Brinkman-Morel (ABM) state~\cite{abm1,abm2} which is the chiral $p$-wave pairing state with the orbital angular momentum $L_z=1$ and spin $S_z=0$ of the Cooper pair. In a thin film with a thickness much shorter than the dipole coherence length $\sim 10\mu {\rm m}$, strong pair breaking effect on surface restricts the orbital motion of chiral Cooper pairs into a two-dimensional plane, where the topological Fermi arc appears at the edge of the system~\cite{tsutsumi:2010b,tsutsumi:2011b,silaevJETP14, silaev:2012}. Recently, Ikegami {\it et al}.~\cite{ikegami1,ikegami2} directly observed the chirality of the Cooper pairs through intrinsic Magnus force~\cite{salmelin1,salmelin2} acting on injected electrons in the surface of $^3$He-A. The definite chirality of the Cooper pairs generates the skew scattering of quasiparticles at injected electrons. The other candidates of Weyl SCs are the uranium compounds URu$_2$Si$_2$~\cite{yamashita,schemmPRB15}, {UCoGe~\cite{mineev}, U$_{1-x}$Th$_x$Be$_{13}$~\cite{shimizu,mizushimaPRB18,machidaJPSJ18},} and UPt$_3$~\cite{joynt,schemm14}. In URu$_2$Si$_2$, Yamashita {\it et al}.~\cite{yamashita} observed the giant Nernst effect which is associated with the skew scattering of electrons due to the chilarity of superconducting gap $\Delta ({\bm k})\propto \hat{k}_z(\hat{k}_x+i\hat{k}_y)$~\cite{sumiyoshi}. {Schemm {\it et al}. reported the broken time-reversal symmetry in URu$_2$Si$_2$ through the polar Kerr effect~\cite{schemmPRB15}.} UPt$_3$ has fascinated many physicists since the first discovery of superconductivity, because it possesses similar properties with $^3$He as a spin-triplet superconductor. Among many proposed scenarios, the $E_{2u}$ and $E_{1u}$ states are most promising and competing scenarios~\cite{sauls94,tsutsumiJPSJ12-2,machidaPRL12,izawa}. The time-reversal broken $E_{2u}$ states, $\Delta ({\bm k})=i{\bm \sigma}\cdot \hat{\bm c}\sigma _y(\hat{k}_a+i\hat{k}_b)^{2}\hat{k}_c$, which assume a strong spin-orbit coupling, may possess two separated Fermi arcs terminated to Weyl points on a surface parallel to the $c$-axis~\cite{goswami14,mizushimaJPSJ15}. Both the time-reversal broken and invariant pairings are possible in the $E_{1u}$ scenario, where surface Fermi arc in the time-reversal-invariant $E_{1u}$ state, $\Delta ({\bm k})=i{\bm \sigma}\cdot (\hat{\bm a}\hat{k}_b+\hat{\bm b}\hat{k}_a)(5\hat{k}^2_c-1)\sigma _y$ or $i{\bm \sigma}\cdot (\hat{\bm b}\hat{k}_a+\hat{\bm c}\hat{k}_b)(5\hat{k}^2_c-1)\sigma _y$, is protected by an order-two magnetic point group symmetry~\cite{tsutsumiJPSJ13,mizushimaPRB14}. Although both the pairing states have surface Fermi arcs as shown in Figs.~\ref{fig:weyl}(b) and \ref{fig:weyl}(c), only the time-reversal invariant $E_{1u}$ state has anisotropic magnetic response associated with Majorana Ising spins~\cite{tsutsumiJPSJ13,mizushimaPRB14,mizushimaJPSJ15}.

\begin{figure}[!t]
\centering\includegraphics[width=4.8in]{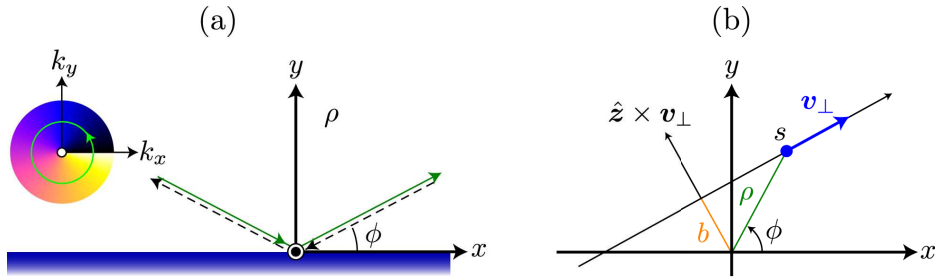}
\caption{Quasiparticle trajectories for surface ABSs (a) and vortex ABSs (b). The color map in (a) shows the phase profile of the chiral $p_x+ip_y$ pairing state in the momentum space. In (b), the vortex center is located at $x=y=0$.
}
\label{fig:trajectory}
\end{figure}

Let us now revisit low-lying surface bound states of Weyl SCs from the viewpoint of the Jackiw-Rebbi's index theorem in the Andreev equation. Consider a specular reflection on the $xz$ surface which is parallel to the nodal direction of the pairing in Eq.~\eqref{eq:deltaell}, and take the coordinate along the quasiclassical trajectory $\rho \equiv y/v_{y}$ for incoming quasiparticle and $\rho \equiv -y/v_{y}$ for outgoing one, where $v_{y}$ denotes the Fermi velocity projected onto the $y$ axis (see Fig.~\ref{fig:trajectory}(a)). The Andreev equation \eqref{eq:a} with the superconducting gap \eqref{eq:deltaell} is then mapped onto the one-dimensional Dirac equation \eqref{eq:dirac} along the trajectory $\rho$, where the effective mass is expressed as $m(\rho) = \Delta _0{\rm e}^{-i\tau _z\vartheta(\rho)}$~\cite{mizushimaJPCM15}. The phase $\vartheta(\rho)$ is given as $\vartheta(\rho)=\ell \phi _{\bm k}$ for $\rho > 0$ and $\vartheta(\rho)=-\ell \phi _{\bm k}$ for $\rho < 0$. This is equivalent to a superconducting junction system with the relative phase difference $2\ell\phi _{\bm k} \in [0,2\pi]$ and thus the ABSs come along the energy dispersion $E(\phi _{\bm k})=\Delta _0 \cos(\ell \phi _{\bm k})$ which is dispersive on $k_x$ but flat band along the nodal direction $k_z$. According to the condition in Eq.~\eqref{eq:pi}, the chiral $\ell$-wave pairing state has $\ell$ gapless points at~\cite{mizushimaJPCM15}
\beq
\hat{k}_x = \sin\!\theta _{\bm k}\cos\left[ \left( n - \frac{1}{2}\right) \frac{\pi}{ \ell }\right] ,
\quad n = 1, 2, \cdots, \ell ,
\label{eq:kx}
\eeq
which is insensitive to the detailed spatial profile of $\Delta _0$ around the surface. For a specular surface perpendicular to the $y$ axis, the $\ell$ Fermi arcs appear in the momentum space $(k_x,k_z)$ projected onto the surface, which are terminated at the projection of pairwise Weyl points at $(k_x,k_z)=(0,+ k_{\rm F})$ and $(0,- k_{\rm F})$. Owing to the spontaneous breaking of the time-reversal symmetry, the dispersion of surface ABSs must be asymmetric in $k_x$, i.e., $E(k_x,k_z) = -E(-k_x,k_z)$, which is called the chiral edge state. 

Since the negative energy part of the branch is occupied in the ground state, the surface ABSs carry the spontaneous mass current in equilibrium~\cite{stone,sauls:2011,tsutsumi:2012,tsutsumiJPSJ12,mizushimaPRA10,mizushimaJPSJ15}. In the case of the chiral $p$-wave pairing ($\ell=1$), chiral edge ABSs along the wall of a cylindrical container carry the total angular momentum, $L^{\rm ABS}_z=N\hbar$, where $N$ denotes the number of electrons. However, the emergence of the low-lying ABSs deviates the continuum states with $|E|>\Delta _0$ from those in the bulk without surfaces. In other words, the existence of the mass domain wall in Eq.~\eqref{eq:dirac} gives rise to the phase shift of the continuum states with $|E|>|m(\pm\infty)|$, in addition to the emergence of low-lying bound states. The significant contribution of the continuum state to the net edge mass current was first pointed out by Stone and Roy~\cite{stone}. The modified continuum states are found to carry the net angular momentum $L^{\rm cont}_z=-N\hbar/2$ and the total angular momentum in the chiral $p$-wave pairing is $L_z=L^{\rm ABS}_z+L^{\rm cont}_z=N\hbar/2$. Hence, the emergence of surface ABSs in unconventional SFs/SCs simultaneously generates the ``backaction'' in the vacuum state.

\subsection{Vortex Andreev bound states}

The condition in Eq.~\eqref{eq:piphase} can also be applied to zero energy states bound to a vortex core of SCs/SFs. The chiral $\ell$-wave superconducting gap having a single vortex is given as
\beq
\Delta ({\bm k},{\bm r}) = {\rm e}^{i\kappa\phi}\Delta (\rho)(\hat{k}_x+i\hat{k}_y)^{\ell},
\label{eq:Dpvoretx}
\eeq
where we introduce the cylindrical coordinate, ${\bm r}=(\rho\cos\phi,\rho\sin\phi,z)$. Here, we consider a straight and axisymmetric vortex line along the $z$ axis and $\kappa \!\in\! \mathbb{Z}$ denotes the vorticity. The superconducting gap vanishes at the vortex center where the phase is not well-defined, and recovers to the bulk value $\Delta _0=\Delta (\rho\rightarrow\infty)$ within the length scale of $\xi$. Since the phase continuously changes from $0$ to $2\kappa\pi$ along the circumference enclosing the core, quasiparticles traveling along a trajectory across the vortex center experience the phase shift of $\kappa\pi$. Hence, the low-energy effective theory can be mapped onto the Andreev equation in a SNS junction or the one-dimensional Dirac equation with a mass domain wall. Thus, singular vortices in SCs/SFs are accompanied by low-lying quasiparticles states, i.e., vortex ABSs.

As shown in Fig.~\ref{fig:trajectory}(b), we introduce a quasiclassical trajectory along $\hat{\bm v}_{\perp}$ parameterized as ${\bm r}=b(\hat{\bm z}\times \hat{\bm v}_{\perp}) + s\hat{\bm v}_{\perp}$ with an impact parameter $b$ and the ``coordinate'' $s\in(-\infty,+\infty)$~\cite{kopnin}. The unit vector $\hat{\bm v}_{\perp}$ denotes the orientation of the Fermi velocity projected onto the $xy$ plane, $\hat{\bm v}_{\perp}=(\cos\alpha,\sin\alpha)$. The parameters $(b,s)$ are associated with the angles $\phi$ and $\alpha$ as $b=\rho\sin(\phi-\alpha)$ and $s=\rho\cos(\phi-\alpha)$. The azimuthal angle $\phi$ is parameterized as $\phi(s)=\alpha + \tan^{-1}(b/s)$, which changes from $\phi(\infty)=\alpha$ to $\phi(-\infty)=\pi+\alpha$. For a single vortex state with Eq.~\eqref{eq:Dpvoretx}, then, the Andreev equation \eqref{eq:a} is recast into the one-dimensional Dirac equation, 
$\left[ -iv \partial _s\hat{\tau}_x + m(s)\hat{\tau}_z \right]\tilde{\bm \varphi}_E(s) = E \tilde{\bm \varphi}_E(s)$, 
by employing the gauge transformation of the quasiparticle wave function $\tilde{\bm \varphi}\mapsto {\rm e}^{i(\kappa+{\ell})\tau _z\alpha/2}\tilde{\bm \varphi}$ ($v\equiv {\bm v}_{\rm F}\cdot\hat{\bm v}_{\perp}$). The effective mass is expressed in terms of the vortex phase winding $\kappa\phi(s)$ as
\beq
m(s) = \Delta (\rho) {\rm e}^{i\kappa(\phi(s)-\alpha)\tau _z}.
\eeq
Since the phase factor of $m(s)$ is reduced to ${\rm sgn}(s)$ at $b=0$, quasiparticles across the vortex center with odd $\kappa$ experience the mass domain wall, $m(s)=-m(-s)$, which satisfies the condition \eqref{eq:piphase}. The correction of a small impact parameter $b\ll \xi$ deviates the relative phase from $\pi$. This indicates that the vortex bound states have dispersion with respect to the impact parameter $b$ as $E(b,k_z) \propto -b$, where $k_z$ denotes the axial momentum along the vortex line. The wave function of the vortex bound states is well localized in the region of $\rho \sim b$ and has the azimuthal momentum $k_{\rm F}$ for small $k_z$. From the quasiclassical viewpoint, therefore, the impact parameter is associated with the angular momentum of the vortex bound states as ${\bm L}={\bm b}\times{\bm k}_{\rm F}$. As a result, the quasiclassical dispersion is linear in $L$ and dispersionless in $k_z$ as $E(L,k_z)\approx -\omega _0 L$, where $\omega _0$ denotes the energy scale of the vortex bound states and is determined by the detailed structure of $\Delta (\rho)$. However, {the higher-order corrections to the quasiclassical equation, including particle-hole asymmetry, discretizes $L$ and} may lift the zero energy solution. Since $L$ is the canonically conjugate variable of $\alpha$, the effective Hamiltonian for the vortex bound states is given as the one-dimensional chiral Hamiltonian $\mathcal{H}_{\rm vortex} \approx i\omega _0\partial _{\alpha}$. Volovik~\cite{volovik} found that owing to the single valuedness of the wave function, the discretized level of the vortex bound states remains as zero energy when the net parity of the vorticity $\kappa$ and chirality $\ell$ is even,
\beq
(-1)^{\kappa+\ell}=+1.
\label{eq:parity}
\eeq
This includes chiral $p$-wave SCs with odd vorticity. For net odd parity, $(-1)^{\kappa+\ell}=-1$, the quantum correction prohibits the presence of zero energy vortex bound states. Tewari {\it et al.}~\cite{tewariPRL07} directly demonstrated that for chiral $p$-wave pairing, the Hamiltonian in the sector of the zero energy state is mapped onto the Majorana (or Dirac) Hamiltonian with an effective mass that satisfies the condition \eqref{eq:piphase}, when the vorticity is odd. The index theorem can be extended to the vortex state of noncentrosymmetric superconductors that may possess the mixing of spin-triplet and spin-singlet pairings~\cite{satoPRB09-2}. 

In $s$-wave SCs ($\ell=0$), the direct calculation of the BdG equation within $k_{\rm F}\xi \gg 1$ shows $E_{m} \!=\! - ( m - 1/2) \omega _0$ ($m\in\mathbb{Z}$), where the zero energy state is always absent. The discretized core levels, which are called the Caroli-de Gennes-Matricon (CdGM) state~\cite{cdgm}, were clearly observed in the spin-singlet SC YNi$_2$B$_2$C by using scanning tunneling spectroscopy~\cite{kaneko}. As the quantum regime ($\xi \sim k^{-1}_{\rm F}$) is approached, the level spacing $\omega _0 \approx \Delta^2_0/E_{\rm F}$ becomes large, leading to the quantum depletion of the particle density around the core~\cite{hayashiPRL98,hayashiJPSJ98}. Pronounced depletions in the particle density were experimentally observed in rotating Fermi gases with an $s$-wave resonance \cite{mit} as a hallmark of superfluidity in the BCS-BEC crossover regime. The depletion also indicates the absence of vortex ABSs in the core. 

Similarly with solitonic states in FFLO SCs, the CdGM states can accumulate Pauli paramagnetic moments in the vortex core when magnetic Zeeman field or population imbalance of two spin states is present~\cite{takahashiPRL06,ichiokaPRB07}. A strong Pauli-paramagnetic effect in spin-singlet SCs drastically changes thermodynamics and transport properties in low temperatures from those in orbital-effect-dominant SCs~\cite{ichiokaPRB07}. For instance, the field dependence of low-temperature heat capacity shows a concave curve $C(H)\propto (H/H_{\rm c2})^{n}$ ($n>1$) when the Pauli paramagnetic effect is dominant over the orbital depairing effect. The concave curves have been observed in heavy fermion SCs CeCoIn$_5$~\cite{115}, UBe$_{13}$~\cite{ube13,shimizu}, CeCu$_2$Si$_2$~\cite{kittaka}. The similar field dependence was also observed in Sr$_2$RuO$_4$~\cite{deguchi}. It is also worth mentioning that, if vortices and FFLO nodal planes coexist, the nontrivial topological structure of superconducting gap gives rise to the absence of low-lying ABSs at the crossing point~\cite{mizushimaPRL05,ichiokaPRB07-2}. This is attributed to the cancelation of the $\pi$-phase shift on a vortex line by an FFLO nodal plane. At this crossing point, the absence of low-lying ABSs results in the depletion of the local magnetization density. We also notice that in the quantum limit with $\Delta^2/E_{\rm F}\sim O(\Delta)$, the spatial oscillation of the magnetization density inside the core manifests the energy level and vorticity of the superconducting gap and thus can be utilized for a fingerprint of core-level spectroscopy~\cite{takahashiPRL06,suzukiPRA08}. {Furthermore, Eschrig {\it et al.} found that vortex core dynamics in unconventional SCs closely reflects features of the vortex ABSs~\cite{eschrigPRB99,saulsNJP99}.}


The discretized energy levels of vortex ABSs in the chiral $p$-wave pairing ($\ell =1$) is obtained as~\cite{mizushimaPRA10} 
\begin{eqnarray}
E_{m} = - \left( m - \frac{\kappa + \ell}{2}\right) \omega _0 ,
\label{eq:cdgm}
\end{eqnarray}
when the chemical potential is positive ($\mu>0$). In contrast to the $s$-wave case, the zero energy state always exists when the condition of the net parity in Eq.~\eqref{eq:parity} is satisfied. The zero energy state realized in this system is protected by the $\mathbb{Z}_2$ topological invariant, $\nu$, which is the parity of the second Chern number ${\rm Ch}_2=\kappa \ell$~\cite{teoPRB10}. The zero energy vortex ABS exists if the $\mathbb{Z}_2$ number is $\nu=(-1)^{{\rm Ch}_2}=-1$, which is consistent with \eqref{eq:parity}. The $\mathbb{Z}_2$ number indicates that the zero energy state is fragile against the quasiparticle tunneling to the neighboring vortex. Indeed, the intervortex tunneling splits the zero energy states to $\pm \sin(k_{\rm F}D)e^{-D/\xi}$~\cite{mizushimaPRA10-2,chengPRL09}, {depending on the intervorterx distance $D$.} The sinusoidal and exponential factors represent the quantum oscillation of the zero energy wave functions and localization at the vortex core, respectively. This is contrast to vortices realized in the ($2+1$)-dimensional Dirac equation, where the number of zero energy states is guaranteed by the topological $\mathbb{Z}$ number~\cite{weinberg,rossi}. The zero energy states protected by the $\mathbb{Z}$ number appear at the interface of three-dimensional topological insulator (AII) and conventional superconductor with a fine-tuned chemical potential~\cite{chengPRB10,chiuPRB15}. When the chemical potentail becomes negative, however, the zero energy vortex ABS disappears and the topological number is trivial, regardless of the vorticity and chirality. Hence, similarly with surface ABSs, the topological phase transition occurs at $\mu=0$, where the bulk excitation becomes gapless. 


\section{Andreev bound states meet symmetry}
\label{sec:symmetry}

In the previous section, we have shown a unified description for surface and vortex ABSs and nontrivial topological properties behind them. To extract multifaceted properties of ABSs and capture the essential roles of discrete symmetries, we here focus on the superfluid $^3$He, where various nontrivial phenomena associated with the interplay of topology and symmetry can be realized~\cite{mizushimaJPSJ15}. We start to introduce Majorana fermions that can be regarded as a special kind of ABSs, where we mention non-Abelian anyonic behaviors of mirror-symmetry-protected Majorana fermions in spinful odd parity SCs/SFs, such as $^3$He-A~\cite{sato14,ueno}. Then, we give an overview on symmetry-protected topological superfluidity realized in $^3$He-B confined to a slab geometry. This offers a unique platform to study the intertwining of mass acquisition of surface ABSs and spontaneous symmetry breaking that protects the nontrivial topology of the bulk, resulting in the emergence of ``topological quantum criticality''~\cite{mizushimaJPSJ15,mizushimaPRL12}. We also clarify anomalous magnetic response of surface ABSs in the vicinity of the topological quantum criticality on the basis of odd frequency pairing which is another property of ABSs. 

\subsection{Andreev bound states as Majorana fermions}
\label{sec:majorana}

In relativistic field theory, self-conjugate solutions of the Dirac equation are called Majorana fermions. They are represented by the quantized field ${\bm \Psi}$ that satisfies the constraint of the self-charge-conjugation in Eq.~\eqref{eq:majorana}. The constraint is satisfied only when the antiunitary particle-hole (or charge conjugation) operator $\mathcal{C}$ obeys 
$\mathcal{C}^2 = +1$, corresponding to odd parity SCs/SFs. In even parity SCs/SFs, however the pair potential $\Delta ({\bm k})$ is always invariant under the spin rotation, and the particle-hole operator is given by $\mathcal{C}^2\!=\! -1 $. ABSs emergent to even parity SCs/SFs can not satisfy Eq.~\eqref{eq:majorana} and thus can not be Majorana fermions. The presence of a strong spin-orbit interaction may mix the spin-singlet and triplet pairings, which enables even spin singlet superconductors to host Majorana fermions~\cite{satoPL03,satoPRB09-2,satoPRL09,satoPRB10-2}.

The fermionic field ${\bm \Psi}$ is in general expanded in terms of energy eigenstates as in Eq.~\eqref{eq:psi}. If there exist $n$ zero-energy states ${\bm \varphi}^{(a)}_{E=0}({\bm r})$ ($a=1,\cdots,n$), one can rewrite ${\bm \Psi}$ as 
\beq
{\bm \Psi}({\bm r}) = \sum^n_{a=1} {\bm \varphi}^{(a)}_{0}({\bm r}) \gamma^{(a)}
+ \sum _{E>0} \left[ 
{\bm \varphi}_E({\bm r}) \eta _E + \mathcal{C}{\bm \varphi}_E({\bm r}) \eta^{\dag}_E
\right],
\label{eq:PsiM}
\eeq
where we have used $\gamma^{(a)}$, instead of $\eta_{E=0}$, in order to distinguish these zero modes. As explicitly shown in Eq.~\eqref{eq:diracwf}, the zero energy states are composed of equal contributions from the particle-like and hole-like components of quasiparticles. This is the rigorous result guaranteed by the particle-hole symmetry of the zero energy state, ${\bm \varphi}_0({\bm r})=\mathcal{C}{\bm \varphi}_0({\bm r})$, where the BdG Hamiltonian in Eq.~\eqref{eq:bdgH} holds $\mathcal{C}\mathcal{H}({\bm k},{\bm r})\mathcal{C}^{-1}=-\mathcal{H}(-{\bm k},{\bm r})$. The self-conjugate constraint in Eq.~\eqref{eq:majorana} imposes the following relations, 
\beq
\gamma^{(a)} = \gamma^{(a)\dag}.
\label{eq:majo}
\eeq
The zero modes satisfy the self-conjugate constraint, which are called as Majorana zero modes. Since surface and zero-energy vortex ABSs in spin-triplet SCs always satisfy the self-conjugate condition, Majorana fermions can be regarded as a special property of ABSs. When there is only a single zero mode ($n=1$), the local density operator defined in the particle-hole space is identically zero, 
\beq
\rho ({\bm r}) \equiv {\bm \Psi}^{\dag}({\bm r})\tau _z {\bm \Psi}({\bm r})/2 = 0. 
\label{eq:rho}
\eeq
This indicates that an isolated Majorana zero mode can not be coupled to the local density fluctuation and thus is very robust against non-magnetic impurities.

The relation in Eq.~\eqref{eq:majo} also gives rise to a significant feature inherent to Majorana field. For instance, 
In the case of $n=2$, the minimal representation of the algebra in Eq.~\eqref{eq:majo} is two-dimensional, which is built up by defining {\it complex fermion} operators $c$ and $c^{\dag}$ as
\beq
c = \frac{1}{\sqrt{2}} (\gamma^{(1)}+i\gamma^{(2)}), \hspace{3mm}
c^{\dag} = \frac{1}{\sqrt{2}} (\gamma^{(1)}-i\gamma^{(2)}).
\label{eq:complex}
\eeq 
These operators obey anti-commutation relations, $\{c,c^{\dag}\}=1$ and $\{c,c\}=\{c^{\dag},c^{\dag}\}=0$. The two degenerate vacua $|\pm \rangle$ are defined as the eigenstate of the fermion parity, where $| + \rangle$ ($| - \rangle$) has the even (odd) fermion parity. The eigenstate of $\gamma^{(a)}$ is a superposed state of opposite fermion parities, while the complex fermion preserves the fermion parity and thus can be a physical state~\cite{semenoff}. Equations \eqref{eq:majo} and \eqref{eq:complex} bring two novel quantum phenomena: (i) non-local correlation~\cite{semenoff} and (ii) non-Abelian statistics~\cite{ivanovPRL01}. These two phenomena require Majorana zero modes to be spatially separated and well isolated from other quasiparticle states with higher energies. As mentioned in the previous section, a quantum vortex with a singular core in chiral $p$-wave SCs/SFs can host topologically protected Majorana zero modes. Ivanov~\cite{ivanovPRL01} clarified that the low-energy physics on half-quantum vortices in spinful chiral $p$-wave SCs/SFs, e.g., $^3$He-A, is describable with singular vortices in a spin-polarized system, and the Majorana zero modes concomitant with vortices behave as non-Abelian anyons. For {\it spinless} chiral $p$-wave SCs/SFs with $2N$ vortices, $N$ complex fermions constructed from the $2N$ Majorana zero modes give rise to $2^{N-1}$-fold degeneracy of ground states with preserving fermion parity. The representations of the braiding operation of vortices are obtained as a discrete set of the unitary group which manipulates the occupation of complex fermions. Since Majorana zero modes are topologically protected against quantum decoherence, they offer a promising platform to realize topological quantum computation~\cite{ivanovPRL01,nayakRMP08}.

Although spin-polarized Majorana zero modes behave as non-Abelian anyons, the realization of half-quantum vortices remains as an experimentally challenging task in both SF $^3$He-A thin film and spin-triplet SCs such as Sr$_2$RuO$_4$~\cite{chungPRL07,kawakamiPRB09,vakaryukPRL09,kawakamiJPSJ10,kawakamiJPSJ11,kondoJPSJ12,nakaharaPRB14,yamashitaPRL08}. The key to implement non-Abelian statistics of braiding vortices is a non-local pair of Majorana zero modes that form the complex fermion~\eqref{eq:complex}. Spinful Majorana zero modes can form the complex fermion as a local pair, $c=(\gamma^{(\uparrow)}+i\gamma ^{(\downarrow)})$. However, it is recently recognized that mirror reflection symmetry may assist non-Abelian statistics of spinful Majorana zero modes realized in integer vortices of spin-triplet chiral $p$-wave SFs/SCs~\cite{ueno,sato14,fangPRL14,liuprx14}.

Let us assume that the normal state holds the mirror symmetry with respect to the $xy$ plane, $M_{xy} \varepsilon ({\bm k}) M^{\dag}_{xy} = \varepsilon (k_x,k_y,-k_z)$, where $M_{xy}=i\sigma _z$ flips the spin ${\bm \sigma}$ and momentum ${\bm k}$ to $(-\sigma _x,-\sigma _y,\sigma _z)$ and $(k_x,k_y,-k_z)$. When $\Delta ({\bm k})$ has a definite parity under $M_{xy}$ as~\cite{ueno,sato14}
\beq
M_{xy}\Delta ({\bm k})M^{\rm t}_{xy} = \eta \Delta (k_x,k_y,-k_z), \quad \eta = \pm
\label{eq:dmirror}
\eeq 
the BdG Hamiltonian at $k_z=0$ is invariant under the mirror reflection,
$[ {\mathcal{M}}^{\eta}, \mathcal{H}(k_x,k_y,0) ] = 0$, where $\mathcal{M}^{\eta} = {\rm diag}(M_{xy},\eta M^{\ast}_{xy})$.
Therefore, in the mirror invariant plane, ${\bm k}=(k_x,k_y,0)$, spinful quasiparticles are separated into the subsectors constructed from the eigenstates of the mirror operator. In each mirror subsector, there exists a single zero mode bound to the core of an integer vortex in chiral $p$-wave SFs/SCs, which is protected by the topological $\mathbb{Z}_2$ number. The properties of zero modes are, however, categorized into two classes, depending on the parity in Eq.~\eqref{eq:dmirror}. When the mirror operator satisfies the condition, 
\beq
\left\{ \mathcal{C}, \mathcal{M}^{\eta}\right\} = 0,
\label{eq:mirrorMF}
\eeq
the particle-hole symmetry exists in each mirror subsector and the existence of the Majorana fermion is ensured by the mirror reflection symmetry. In spin-triplet SCs/SFs, the physical meaning of the condition \eqref{eq:mirrorMF} is associated with the orientation of the ${\bm d}$-vector. Even though integer quantum vortices are accompanied by spinful Majorana fermions, the mirror symmetry protects multiple Majorana fermions as non-Abelian anyons unless a perturbation that breaks mirror reflection symmetry is present. 

If the mirror operator violates the condition in Eq.~\eqref{eq:mirrorMF}, the particle-hole exchange maps a quasiparticle state to that in the different mirror subsector. This implies the absence of the particle-hole symmetry in each mirror subsector. The topological properties and nature of the zero-energy vortex ABS are categorized to the class A in the Altland-Zirnbauer symmetry classes, where the zero energy states behave as Dirac fermions~\cite{ueno,sato14}.

\subsection{Helical Majorana fermions: Surface Andreev bound states in $^3$He-B}

Although the Weyl points are protected by the first Chern number and are responsible for the existence of the Fermi arc, it is known that the topological stability of point nodes is subject to discrete symmetries, such as the time-reversal symmetry. In particular, point nodes in time-reversal invariant SFs and SCs are no longer protected by the topological invariant when additional discrete symmetries are absent. We here discuss surface ABSs in the B and planar phases of the SF $^3$He, as a paradigm to capture an essence of additional discrete symmetries. The additional discrete symmetries, such as magnetic point group symmetry, enrich the characteristics of ABSs, including Majorana Ising spins~\cite{mizushimaPRL12,mizushimaJPCM15,mizushimaJPSJ15}. 

The $^3$He atom is a neutral atom with nuclear spin $1/2$ and zero electron spin, and the system remains liquid phase down to zero temperatures. The quantum liquid is well describable with the strongly correlated Fermi liquid theory which holds the huge symmetry group,
$G = {\rm SO}(3)_{\bm L} \times {\rm SO}(3)_{\bm S} \times {\rm U}(1)_{\phi} \times {\rm T} $. 
This contains three-dimensional rotations in coordinate and spin spaces (${\rm SO}(3)_{\bm L}$ and ${\rm SO}(3)_{\bm S}$), the global phase transformation group (${\rm U}(1)_{\phi}$), and the time-reversal symmetry (${\rm T}$). Among possible broken symmetries of $G$, the phase which holds the maximal symmetry group, 
${H}_{\rm B} = {\rm SO}(3)_{{\bm L}+{\bm S}} \times {\rm T}$, is known as the B-phase or the Balian-Werthamer (BW) phase. The BW state is the spontaneous breaking phase of the spin-orbit symmetry which is characterized by the broken symmetry group 
$\mathcal{R}_{\rm B}=G/H_{\rm B} = {\rm SO}(3)_{{\bm L}-{\bm S}} \times {\rm U}(1)_{\phi}$, 
where ${\rm SO}(3)_{{\bm L}-{\bm S}} $ is the relative rotation of the spin and orbital spaces. The generic form of the BW state that characterizes the order parameter manifold $\mathcal{R}_{\rm B}$ is obtained as
\beq
d_{\mu}({\bm k}) = \Delta _{\rm B} R_{\mu\nu}(\hat{\bm n},\varphi) \hat{k}_{\nu},
\label{eq:dvecB}
\eeq
where $d_{\mu}({\bm k})={\rm tr}[-i\sigma _y\sigma _{\mu}\Delta ({\bm k})]/2$. The rotation matrix $R_{\mu\nu}$ is associated with ${\rm SO}(3)_{{\bm L}-{\bm S}}$ and we omit the ${\rm U}(1)$ phase. In the bulk BW state without a magnetic field, the angle $\varphi$ is fixed to the so-called Leggett angle, 
$\varphi _{\rm L} = \cos^{-1}\left( -\frac{1}{4}\right)$. The $\hat{\bm n}$-vector is affected by confinement and magnetic field. The bulk BW state has a fully gapped quasiparticle spectrum as $E({\bm k}) = \pm \sqrt{[\varepsilon ({\bm k})]^2 + \Delta^2_{\rm B}}$.

{\it Surface ABS and its helicity.}--- Let us now derive gapless states bound to the surface of the BW state, by mapping the Andreev equation \eqref{eq:a} to Eq.~\eqref{eq:dirac}. Here, we set a specular surface to be normal to the $\hat{\bm z}$-axis and the region $z>0$ is occupied by the $^3$He-B. The unitary matrix $S(\phi_{\bm k}) \!\equiv\! (\sigma _x+\sigma _z)e^{i\vartheta\sigma _z}/\sqrt{2}$ with $\vartheta = \frac{\phi _{\bm k}}{2} - \frac{\pi}{4}$ maps the Andreev equation \eqref{eq:a} with Eq.~\eqref{eq:dvecB} onto the time-reversal-symmetric pair of chiral $p$-wave SFs~\cite{mizushimaPRB12,mizushimaJPCM15,nagaiJPSJ08}
\begin{align}
{\Delta}(\hat{\bm k},{\bm r}) 
= U(\hat{\bm n},\varphi) S(\phi_{\bm k})
\left( 
\begin{array}{cc}
\Delta _{\rm B}e^{i\theta _{\bm k}} & 0 \\
0 & -\Delta _{\rm B} e^{-i\theta _{\bm k}}
\end{array}
\right)
S^{\rm t}(\phi_{\bm k})U^{\rm t}(\hat{\bm n},\varphi),
\label{eq:unitaryB} 
\end{align}
where $U(\hat{\bm n},\varphi)$ is the ${\rm SU}(2)$ spin rotation associated with $R_{\mu\nu} $ and we set ${\bm k}_{{\rm F}} \!=\! k_{\rm F}(\cos\!\phi _{\bm k}\sin\!\theta _{\bm k},\sin\!\phi _{\bm k}\sin\!\theta _{\bm k},\cos \!\theta _{\bm k})$. 
Hence, the general concept based on Eq.~\eqref{eq:dirac} can be extended to surface ABSs of $^3$He-B, which ensures the existence of the zero energy ABSs on the surface of $^3$He-B. The bound state solution with $|E({\bm k}_{\parallel})| \!\le\! \Delta _{\rm B}$ has the gapless dispersion as
\beq
E({\bm k}_{\parallel}) =  \frac{\Delta _{\rm B}}{k_{\rm F}} |{\bm k}_{\parallel}|,
\label{eq:E0a}
\eeq 
where ${\bm k}_{\parallel} \!=\! (k_x,k_y)$ denotes the momentum parallel to the surface. The wave function for the positive energy branch of the surface ABS in Eq.~\eqref{eq:E0a} is given as
\beq
{\bm \varphi}_{{\bm k}_{\parallel}}({\bm r}) = \mathcal{N} e^{i{\bm k}_{\parallel}\cdot{\bm r}_{\parallel}}
f(k_z,z) \mathcal{U}(\hat{\bm n},\varphi) \left[
{\bm \Phi}_+ -e^{i\phi _{\bm k}}{\bm \Phi}_-
\right],
\label{eq:sabswf}
\eeq
where $N$ is the normalization constant. The function $f(k_z,z)\equiv\sin (k_zz)e^{-z/\xi _{\rm B}}$ with $\xi _{\rm B}= v_{\rm F}/m\xi _{\rm B}$ denotes the spatial form of the surface ABSs. The spinors ${\bm \Phi}_+\equiv(1,0,0,i)^{\rm t}$ and ${\bm \Phi}_-\equiv (0,i,1,0)^{\rm t}$ are the eigenstates of the spin operator in the Nambu space, $S_z={\rm diag}(\sigma _z,-\sigma ^{\rm t}_z)/2$. 

We emphasize that the spin structure of the surface ABS reflects the emergence of the spin-orbit coupling through the spontaneous spin-orbit symmetry breaking in the vacuum. The bulk quasiparticle states in the BW state are doubly degenerate as the ${\bm \varphi}_E({\bm k})$ and $\mathcal{TP}{\bm \varphi}_E({\bm k})$, where $\mathcal{T}=i\sigma _y K$ and $\mathcal{P}$ denote the operators of the time-reversal symmetry and inversion symmetry, respectively. They are also the simultaneous eigenstates of the helicity operator in the Nambu space
\beq
h \equiv \left(
\begin{array}{cc}
i\hat{\bm k}\cdot{\bm \sigma} & \\ & -i\hat{\bm k}\cdot{\bm \sigma}^{\rm t} 
\end{array}
\right).
\eeq
Since the $\mathcal{TP}$ operation flips the helicity, the doubly degenerate quasiparticle states in the bulk BW state can be labeled by the eigenvalue of the helicity, $h=\pm 1$. The helicity is canceled out by two bands and the spin current is absent in the bulk. In contrast to the bulk BW state, the whole branch of the surface ABS in Eq.~\eqref{eq:sabswf} has an well-defined helicity $h _{\parallel}=+1$ {\it or} $h _{\parallel}=-1$, where $h_{\parallel}$ denotes the helicity operator projected on the surface. Hence, the surface ABS in $^3$He-B has only a half degrees of freedom in comparison with the bulk quasiparticle states. The well-defined helicity in the whole branch of the surface ABS implies the generation of spontaneous spin current, $J_{\mu\nu}$, which denotes the spin $\nu$ flowing along the $\mu$-direction on the surface. The orientations of the spin and flow depend on the order parameter structure $(\hat{\bm n},\varphi)$ and the nonzero components are given as~\cite{mizushimaJPCM15,tsutsumiJPSJ12}
\beq
J^{\rm spin}_{\mu x}=-R_{\mu x}(\hat{\bm n},\varphi)J^{\rm spin}, \quad
J^{\rm spin}_{\mu y}=R_{\mu y}(\hat{\bm n},\varphi)J^{\rm spin}, \quad
J^{\rm spin}=-\frac{n\kappa\hbar}{6},
\eeq
where $n$ denotes the particle density and we set $\kappa = \hbar/2m$. We notice that similarly with the mass current in Weyl SCs/SFs (see Sec.~2(b)), the spin current is carried by both the ABS with $|E|<\Delta _{\rm B}$ and continuum state with $|E|>\Delta _{\rm B}$, i.e., $J^{\rm spin} = J^{\rm spin}_{\rm ABS} + J^{\rm spin}_{\rm cont}$. However, the spin current $J^{\rm spin}_{\rm ABS}$ carried by the surface ABS flows in the opposite direction of that carried by continuum states. This indicates that the emergence of ABSs in unconventional SCs/SFs generates the ``backaction'' on the vacuum state. The temperature dependence of $J^{\rm spin}$ shows the $T^3$-power behavior in low temperatures. The power law behavior is attributed to thermal excitations in the positive energy branch of surface ABSs and thus different from the exponential behavior of the continuum state~\cite{mizushimaJPCM15,tsutsumiJPSJ12,wuPRB13}.

{\it Topology and Majorana Ising spins.}--- {The superfluid $^3$He-B is a prototype of the topological class DIII~\cite{schnyderPRB08}, where the time-reversal symmetry ensures the nontrivial topological structure in the momentum space. The nontrivial topological invariant guarantees the existence of gapless surface ABSs.} A generic form of a $4\times 4$ hermitian matrix subject to time reversal, particle-hole, and inversion symmetries is expanded in terms of the four Dirac $\gamma$-matrix as 
$\mathcal{H}({\bm k}) = \sum^4_{j=1} {m}_j({\bm k})\gamma _j $,
where we choose $(\gamma _1, \gamma _2, \gamma _3, \gamma _4) \!=\!  
(- \sigma_z \tau _x, - \tau _y,  \sigma_x \tau _x, \tau _z)$.
As a result, the Hamiltonian is generally parametrized with the four dimensional vector $[m_1({\bm k}),m_2({\bm k}),m_3({\bm k}),m_4({\bm k})]$. 
In the above representation of $\gamma$ matrices, the chiral operator $\Gamma$ is written as $\Gamma \!=\!\gamma_5
\!\equiv\!\gamma_1\gamma_2\gamma_3\gamma_4$. Flattening the Hamiltonian to $Q({\bm k})$, where $Q^2({\bm k})=+1$, we introduce the four dimensional spinor $\hat{m}_{j}({\bm k})=m_{j}({\bm k})/|{\bm m}({\bm k})|$ that contains all the informations on $\mathcal{H}$, so it defines a three dimensional sphere $S^3$ with unit radius. The spinor $\hat{\bm m}({\bm k})$ defines the images of the three-dimensional ${\bm k}$-space compactified on $S^3$ onto the target space $S^3$ that characterizes the Hamiltonian $\mathcal{H}$. The nontrivial mapping is represented by the homotopy group $\pi _3(S^3)$ and the winding number is given as
\begin{eqnarray}
w_{\rm 3d}=\int \frac{d{\bm k}}{12\pi^3}
\epsilon_{\mu\nu\eta}\epsilon_{ijkl}
\hat{m}_i({\bm k})
\partial_\mu\hat{m}_j({\bm k})
\partial_\nu\hat{m}_k({\bm k})
\partial_\eta\hat{m}_l({\bm k}). 
\label{eq:w3d}
\end{eqnarray}
This indicates that $w_{\rm 3d}$ counts how many times the spinor $\hat{m}_j({\bm k})$ warps the three dimensional sphere when one sweeps the whole momentum space. For $^3$He-B with Eq.~\eqref{eq:dvecB}, the winding number is evaluated as $w_{\rm 3d}=1$, which protects the existence of the gapless surface ABS with an well-defined helicity. The nontrivial topology of the BW state was pointed out by Schnyder {\it et al}.~\cite{schnyderPRB08}, Roy~\cite{roy08}, Qi {\it et al}.~\cite{qiPRL09}, Volovik~\cite{volovikJETP09v2}, and Sato~\cite{satoPRB09}. 

Let us now clarify the Majorana nature of topologically protected ABSs and discuss their physical consequences, where the role of additional discrete symmetries is emphasized. Within the Andreev approximation ($\Delta _{\rm B}\ll E_{\rm F}$), as in Eq.~\eqref{eq:sabswf}, the whole branch of the surface ABS has equal contributions from the particle and hole components of quasiparticles. The fermionic field operator constructed from the helical ABS obeys the self-charge conjugation relation 
\beq
\left( 
\begin{array}{c}
\psi _{\uparrow}({\bm r}) \\ \psi _{\downarrow}({\bm r})
\end{array}
\right) = i\sigma _{\mu}R_{\mu z}(\hat{\bm n},\varphi) 
\left( 
\begin{array}{c}
\psi^{\dag}_{\downarrow}({\bm r}) \\ - \psi^{\dag}_{\uparrow}({\bm r})
\end{array}
\right),
\eeq
which is called the helical Majorana fermion. This leads to the relation mentioned in Eq.~\eqref{eq:rho} that the local density operator constructed from the surface ABSs are identically zero. Hence, it is expected that the surface ABS is robust against nonmagnetic impurities. In addition, it turns out that the anomalous spin structure associated with the definite helicity generates the characteristic orientation of the local spin density operator as
\beq
S^{({\rm surf})}_{\mu} = R_{\mu z}(\hat{\bm n},\varphi) S^{\rm M}_z,
\label{eq:MIS3}
\eeq
where $S^{\rm M}_z$ is the logical spin operator constructed from the surface ABS for $\hat{\bm n}=\hat{\bm z}$ and $\varphi = 0$. Equation \eqref{eq:MIS3} indicates that the surface helical ABS represented in Eq.~\eqref{eq:sabswf} has a characteristic uniaxial orientation for the magnetic response, which is called the Majorana Ising spin~\cite{chungPRL09,nagatoJPSJ09,shindouPRB10,mizushimaPRL12,mizushimaPRB12}. The uniaxial orientation is determined by the order parameter of the BW state, $(\hat{\bm n},\varphi)$, and thus the surface helical ABS can be coupled to an external magnetic field only when the order parameter satisfies the condition
\beq
\hat{\ell}_z(\hat{\bm n},\varphi) \equiv \hat{h}_{\mu} R_{\mu z}(\hat{\bm n},\varphi) \neq 0,
\label{eq:ell}
\eeq
where $\hat{h}_{\mu}\equiv H_{\mu}/H$ denotes the orientation of the applied magnetic field. In other words, the surface ABS in $^3$He-B remains gapless even in the presence of a magnetic field unless the condition \eqref{eq:ell} is satisfied.

Equation \eqref{eq:ell} thus indicates that the mass acquisition of surface Majorana fermions in $^3$He-B is associated with the configuration of the order parameter $(\hat{\bm n},\varphi)$ and orientation of an applied magnetic field. The effective Hamiltonian for the surface ABS is indeed mapped to the Hamiltonian for helical Majorana fermions as~\cite{mizushimaJPCM15,mizushimaJPSJ15}
\begin{align}
\mathcal{H}_{\rm surf} = \sum _{{\bm k}_{\parallel}} {\psi}^{\rm t}_{\rm M}(-{\bm k}_{\parallel}) \left[
c \left( {\bm k}_{\parallel} \times {\bm \sigma} \right)\cdot\hat{\bm z}
+M(\hat{\bm n},\varphi)\sigma _z
\right] {\psi}_{\rm M}({\bm k}_{\parallel}),
\label{eq:Hsurf2}
\end{align}
where $c = \Delta _{\rm B}/k_{\rm F}$. The Majorana field ${\psi}_{\rm M}$ is associated with the original quantized field for surface states, $\psi$, as  
$\psi({\bm k}) \equiv U(\hat{\bm n},\varphi){\psi}_{\rm M}({\bm k})$, 
which obeys $\{ \psi _{a}, \psi _b \} = \delta _{ab}$. The mass of helical Majorana fermions is parameterized by $\hat{\ell}_z$ as
\beq
M(\hat{\bm n},\varphi) = \frac{\gamma H}{2}\hat{\ell}_z(\hat{\bm n},\varphi),
\label{eq:Mmass}
\eeq 
where $\gamma$ is the gyromagnetic ratio of $^3$He nuclei. By replacing ${\bm k}_{\parallel}$ to $(-i\partial _x,-i\partial _y)$, one can derive the equation of motion for helical Majorana fermions from the effective Hamiltonian as
\beq
\left( -i\gamma^{\mu}\partial _{\mu} + M(\hat{\bm n},\varphi) \right) \psi _{\rm M}(z) = 0,
\eeq
which reduces to the $2+1$-dimensional Majorana equation. Without loss of generality, we set $M/c \rightarrow M$. The $\gamma$-matrices are introduced as $(\gamma^{0},\gamma^1,\gamma^2) = (\sigma_z,i\sigma _x, i\sigma _y)$, which satisfies $\{\gamma ^{\mu},\gamma^{\nu}\}=2g^{\mu\nu}$ with the metric $g^{\mu\nu}=g_{\mu\nu}={\rm diag}(+1,-1,-1)$ ($\mu,\nu=0,1,2$). The manifestation of the helical Majorana fermions is the coupling to the gravitational field through the gravitational instanton term~\cite{wangPRB11,ryuPRB12}, which is responsible for the quantization of thermal Hall conductivity. It is also predicted that the coupling of the Majorana fermions to the gravitational field gives rise to cross correlated responses~\cite{nomuraPRL12}. 

{\it Topological origin of the mass acquisition.}--- Equation \eqref{eq:Mmass} indicates that as long as $\hat{\ell}_z=0$ is satisfied, the surface helical Majorana fermion remain gapless even in the presence of a magnetic field. We notice that although the massless Majorana fermion under no magnetic field is a consequence of the nontrivial topological invariant $w_{\rm 3d}=1$ of the bulk BW state, the topological invariant in Eq.~\eqref{eq:w3d} is not well-defined in the presence of a time-reversal breaking field. It therefore appears that the existence of the massless Majorana fermion under a magnetic field is not a consequence of the nontrivial topology and merely accidental. Contrary to the naive expectation, it is uncovered in Ref.~\cite{mizushimaPRL12} that the massless Majorana fermion under a magnetic field and its mass acquisition is a consequence of the intertwining effect of the hidden symmetry and topology behind the bulk BW state. The key observations are following two-fold: (i) The parameter $\hat{\ell}_z(\hat{\bm n},\varphi)$ or equivalently the mass $M(\hat{\bm n},\varphi)$ in Eq.~\eqref{eq:Mmass} is identified as the order parameter of an order-two magnetic point group symmetry which is referred to as the $P_3$ symmetry. The {\rm Ising order} $\hat{\ell}_z$ stays zero as long as the $P_3$ symmetry is preserved, but the symmetry breaking generates the nonzero Ising order $\hat{\ell}_z\neq 0$. (ii) A topological invariant can be well-defined if the $P_3$ symmetry is held. The topological invariant not only protects the existence of massless Majorana fermions but also guarantees that helical Majorana fermions possess the Ising-like spin anisotropy in Eq.~\eqref{eq:MIS3}. 

To clarify the key observations, let us consider $^3$He-B confined in a slab geometry, where $^3$He is sandwiched by two parallel surfaces and $\hat{\bm z}$ is normal to the two parallel surfaces. The group symmetry subject to the confined $^3$He under a magnetic field is given by 
\beq
G_{\rm slab} = P_2 \times P_3 \times {\rm U}(1)_{\phi}.
\label{eq:g}
\eeq
The $P_3$ ($P_2$) symmetry is an order-two magnetic point group symmetry constructed by the combination of the time-reversal operation $\mathcal{T}=i\sigma _yK$ and joint $\pi$-rotation in spin and orbital spaces (mirror reflection), where the rotation about the surface normal ${\bm H}\rightarrow(-H_x,-H_y,H_z)$ (the mirror reflection in the $xz$ plane ${\bm H}\rightarrow(-H_x,H_y,-H_z)$) is compensated by $\mathcal{T}$ when $H_z=0$ ($H_y=0$). The combined discrete symmetries can be maintained even if each symmetry is explicitly broken. The operator of the $P_3$ symmetry can be constructed with the $\pi$ spin rotation $\mathcal{U}(\pi)\equiv {\rm diag}(U_z(\pi),U^{\ast}_z(\pi))$ as 
$\mathcal{P}_3 \equiv \mathcal{T}\mathcal{U}(\pi)$ with $\mathcal{P}^2_3 = +1$,
where the joint $\pi$-rotation is expressed in terms of the ${\rm SU}(2)$ matrix $U(\hat{\bm z},\pi)$ as
$U_z(\pi) \equiv U(\hat{\bm n},\varphi) U(\hat{\bm z},\pi) U^{\dag}(\hat{\bm n},\varphi)$.
The $\mathcal{P}_3$ operator transforms the BdG Hamiltonian for the BW state as
\begin{align}
\mathcal{P}_3 \mathcal{H}({\bm k})\mathcal{P}^{-1}_3
= \mathcal{H}(k_x,k_y,-k_z)  - \gamma H \hat{\ell}_z \left(
\begin{array}{cc} 
\tilde{\sigma}_z & 0 \\ 0 & - \tilde{\sigma}^{\ast}_z \end{array} \right),
\label{eq:Z2-2}
\end{align}
where we introduce $\tilde{\sigma}_{z} \equiv \sigma _{\mu}R_{\mu z}(\hat{\bm n},\varphi)$. Hence, $\hat{\ell}_z (\hat{\bm n},\varphi)$ can be regarded as the breaking field of the $P_3$ symmetry. This indicates that there are two possible subphases of $G_{\rm slab}$, the $P_3$ preserving phase (${\rm B}_{\rm I}$) and breaking phase (${\rm B}_{\rm II}$). The B$_{\rm I}$ phase maintains $\hat{\ell}_z=0$, while B$_{\rm II}$ has two degenerate ground states characterized by $\hat{\ell}_z> 0$ and $\hat{\ell}_z<0$. The quantity $\hat{\ell}_z$ mentioned above is transformed nontrivially as 
\beq
\mathcal{P}_3:~ \hat{\ell}_z \mapsto -\hat{\ell}_z,
\label{eq:ellz2}
\eeq 
by the $P_3$ operator. Therefore, $\hat{\ell}_z$ can be interpreted as an order parameter of the order-two magnetic point group $P_3$ symmetry, i.e., the Ising order~\cite{mizushimaPRL12,mizushimaJPCM15,mizushimaJPSJ15}. 

\begin{figure}[!t]
\centering\includegraphics[width=5in]{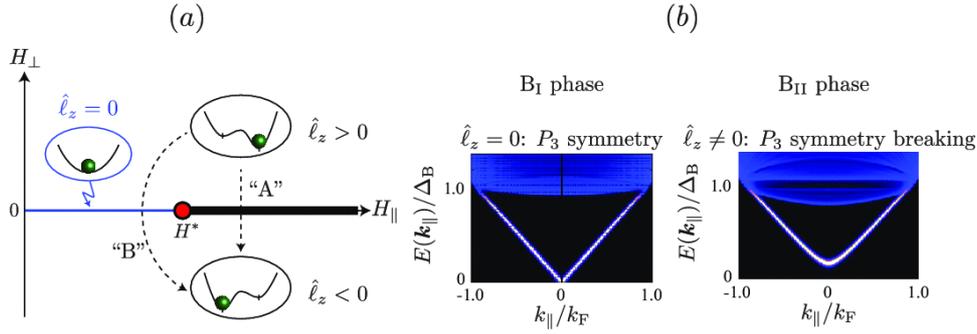}
\caption{(a) Phase diagram of $^3$He-B in a slab geometry under magnetic fields, where $H_{\parallel}$ ($H_{\perp}$) denotes the magnetic field parallel (perpendicular) to the surface and preserves (explicitly breaks) the $P_3$ symmetry. The thin (blue) line corresponds to the symmetry protected topological phase, while the thick line is the non-topological phase with spontaneously breaking the $P_3$ symmetry. The momentum-resolved surface density of states in the corresponding phases is displayed in (b), which shows the dispersion of surface ABSs. 
}
\label{fig:phase}
\end{figure}

By self-consistent calculation based on the microscopic quasiclassical theory, it is demonstrated in Ref.~\cite{mizushimaPRL12} that in $^3$He-B confined to a slab geometry there exists a critical field $H^{\ast}$ beyond which the $P_3$ symmetry is spontaneously broken by increasing a magnetic field parallel to the surface. In Fig.~\ref{fig:phase}(a), we display the phase diagram of $^3$He-B confined to a slab geometry in the presence of a $P_3$ preserving magnetic field $H_{\parallel}$ (parallel to the surface) and breaking field $H_{\perp}$ (perpendicular to the surface). The phase realized in $H<H^{\ast}$ and $H_{\perp} =0$ is identified as the $P_3$ symmetric ${\rm B}_{\rm I}$ phase with $\hat{\ell}_z=0$, while the higher field phase in $H>H^{\ast}$ and $H_{\perp} =0$ (the thick line in Fig.~\ref{fig:phase}(a)) is the ${\rm B}_{\rm II}$ phase characterized by a finite Ising order $\hat{\ell}_z$. The ${\rm B}_{\rm II}$ phase possesses the two degenerate Ising order $+|\hat{\ell}_z|$ and $-|\hat{\ell}_z|$ which are linked to each other by the $P_3$ symmetry operation as in Eq.~\eqref{eq:ellz2}. Within the Ginzburg-Landau theory, $\hat{\ell}_z=0$ for the $P_3$ symmetric ${\rm B}_{\rm I}$ phase is favored by the magnetic dipole interaction acting on nuclear magnetic moment of $^3$He atoms, while the magnetic Zeeman field stabilizes the symmetry broken ${\rm B}_{\rm II}$ phase~\cite{mizushimaJPCM15}. Hence, the critical field is estimated as $H^{\ast}\approx 50{\rm G}$ associated with the dipolar field. In addition to the spontaneous symmetry breaking, as shown in Fig.~\ref{fig:phase}(a), the surface Majorana fermion in the ${\rm B}_{\rm II}$ phase acquires the mass in Eq.~\eqref{eq:Mmass}. 

When $H_{\perp}$ is applied, the $P_3$ symmetry is {\it explicitly} broken and the ground state is characterized by  a definite orientation of the Ising order, $\hat{\ell}_z>0$ {\it or} $\hat{\ell}_z<0$, depending on the orientation of $H_{\perp}$. The resultant phase diagram in Fig.~\ref{fig:phase}(a) has the similar structure with that of the standard Ising model, where the symmetry preserving (breaking) field is replaced to the temperature $T$ (magnetic field $H$). The ${\rm B}_{\rm II}$ phase realized on the thick line of Fig.~\ref{fig:phase}(a) is regarded as the first-order phase transition line between the $\hat{\ell}_z>0$ phase for $H_{\perp}>0$ and $\hat{\ell}_z<0$ for $H_{\perp}<0$. The first order line is terminated at the quantum critical point, which is the critical field $H^{\ast}$. 

We, however, emphasize that the critical field $H^{\ast}$ is also the end point of the topological ${\rm B}_{\rm I}$ phase protected by the $P_3$ symmetry. The key to understand the topological aspect of the ${\rm B}_{\rm I}$ phase is the $P_3$ symmetry in Eq.~\eqref{eq:Z2-2} with $\hat{\ell}_z=0$ and the particle-hole symmetry. Combining these two discrete symmetries, the BdG Hamiltonian of the ${\rm B}_{\rm I}$ phase holds the chiral symmetry, 
\beq
\left\{ \mathcal{H}({\bm k}_{\perp}), \Gamma \right\}=0,
\label{eq:chiral}
\eeq
where ${\bm k}_{\perp}=(0,0,k_z)$ denotes the momentum along the surface normal direction and the chiral operator is defined as  $\Gamma = \mathcal{C}\mathcal{P}_3$. As mentioned above, the nontrivial topological aspect of BdG Hamiltonian for the bulk BW state is understandable as the mapping of the base space ${\bm k}\in S^3$ onto the target space spanned by the four dimensional unit vector $(\hat{m}_1,\hat{m}_2,\hat{m}_3,\hat{m}_4) \in S^3$, when the time-reversal symmetry is preserved. It turns out that the chiral symmetry in Eq.~\eqref{eq:chiral} imposes a contraint on both the base and target spaces. The symmetry reduces to the chiral symmetric momentum path ${\bm k}_{\perp}\in S^1$ and $(\hat{m}_1,\hat{m}_2) \in S^1$. Hence, the topological invariant relevant to the nontrivial mapping $\pi _1(S^1) = \mathbb{Z}$ is given by the one-dimensional winding number~\cite{satoPRB11,mizushimaPRL12,mizushimaJPCM15}
\beq
w_{\rm 1d} = - \frac{1}{4\pi i}\int^{\infty}_{-\infty}dk_z {\rm tr}\left[\Gamma
\mathcal{H}^{-1}({\bm k})\partial _{k_z}\mathcal{H}({\bm k})
\right]_{{\bm k}_{\parallel}={\bm 0}},
\label{eq:w1d}
\eeq
which is evaluated as $w_{\rm 1d}=2$ for the ${\rm B}_{\rm I}$ phase. The chiral symmetry is preserved unless the $P_3$ symmetry is explicitly or spontaneously broken. The bulk-edge correspondence implies the topological protection of the massless nature of surface Majorana fermions in the case of zero Ising order $\hat{\ell}_z=0$. We also notice that the relations on the surface spin in Eqs.~\eqref{eq:MIS3} and \eqref{eq:ell} are a direct consequence of the topological invariant associated with the $P_3$ symmetry, and thus the Majorana Ising spin, $\hat{\bm h}\perp {\bm S}^{\rm surf}$, is a hallmark of the $P_3$ symmetric topological ${\rm B}_{\rm I}$ phase~\cite{mizushimaPRL12}. 

The mass acquisition of Majorana fermions in Eq.~\eqref{eq:Mmass} is directly linked to the topological phase transition concomitant with the spontaneous symmetry breaking of the vacuum. The topological origin of the mass acquisition indicates that the phase diagram around the critical field $H^{\ast}$ essentially differs from that of the standard Ising model. The $\ell _z >0$ and $\ell _z<0$ can not be smoothly connected to each other along the path ``B'' in Fig.~\ref{fig:phase}(a) but the topological phase transition occurs along the path accros $\hat{\ell}_z=0$. The critical field $H^{\ast}$ is therefore regarded as the end point of both the first order phase transition line and topological phase~\cite{mizushimaJPSJ15}. Grover {\it et al.}~\cite{grover} recently predicted the emergence of the supersymmetry at the quantum critical point $H^{\ast}$, which indicates the symmetry under the exchange of fermions (surface Majorana fermions) and bosons (Ising order). 

Wu and Sauls~\cite{wuPRB13} demonstrated that the $P_3$ symmetry is also maintained by the $^3$He-B in the presence of a superfluid flow, where the flow is parallel to the surface. Similarly with the ${\rm B}_{\rm I}$ phase, the topological aspect of $^3$He-B under the superfluid flow is characterized by the nontrivial $w_{\rm 1d}$, which guarantees the existence of gapless Majorana fermions. They proposed that the power-law depletion of the superfluid current is attributed to the thermal excitations of gapless Majorana fermions. 

In addition to the $P_3$ symmetry, there is another order-two magnetic point group symmetry that protects the nontrivial topological properties of $^3$He. This is referred to as the $P_2$ symmetry, where the operator is constructed from the combination of the mirror reflection and time-reversal operation (see Eq.~\eqref{eq:g}). The $P_2$ symmetry is a key to understand the topological aspect of the planar phase whose order parameter is given by $\Delta ({\bm k})=i{\bm \sigma}\sigma _y(\hat{\bm x}\hat{k}_x+\hat{\bm y}\hat{k}_y)$. It is remarkable that although the planar phase is accompanied by pairwise point nodes, they can not be protected by the first Chern number in contrast to Weyl SCs/SFs such as $^3$He-A. In Refs.~\cite{tsutsumiJPSJ13,mizushimaPRB14,mizushimaJPSJ15}, however, it is found that in the planar phase the zero energy flat band terminated by the pairwise point nodes is protected by the $P_2$ symmetry. This is attributed to the fact that the BdG Hamiltonian having the $P_2$ symmetry preserves the chiral symmetry in Eq.~\eqref{eq:chiral} and thus the one-dimensional winding number is well-defined in each one-dimensional momentum path which is perpendicular to the nodal direction ($\hat{k}_z$). This is another mechanism on the stability of the surface Fermi arc which essentially differs from that of Weyl SCs/SFs. We also notice that the concept of the $P_2$ symmetry protected Fermi arc is extended to various superconducting materials, including the time-reversal-invariant $E_{1u}$ scenario of the heavy-fermion UPt$_3$~\cite{tsutsumiJPSJ13} and the $E_u$ state of superconducting topological insulator Cu$_x$Bi$_2$Se$_3$~\cite{sasaki}. Since the Majorana Ising spin is a generic consequence of the chiral symmetry in Eq.~\eqref{eq:chiral}, the surface Fermi arc protected by the $P_2$ symmetry possesses the Ising-like magnetic anisotropy. The Fermi arc is gapped out by only a magnetic field perpendicular to the mirror plane.




\subsection{Andreev bound states as odd-frequency pairing}
\label{sec:odd}

We here give an overview on another property of ABSs in SCs/SFs, that is odd-frequency Cooper pair correlations. In SCs/SFs, the Cooper pair correlation function is defined with the anomalous part of the Matsubara Green's function as 
$
\mathcal{F}_{ab}({\bm r}_1,{\bm r}_2;\tau) = - \langle T_{\tau} \psi _{a}({\bm r}_1,-i\tau) \psi _b ({\bm r}_2,0) 
\rangle$.
This denotes the Cooper pair correlation between the spin $a$ particle at the coordinate ${\bm r}_1$ and time $-i\tau$ and spin $b$ particle at ${\bm r}_2$ and $0$. To classify the possible Cooper pairs, we introduce the following order-two operators, $\hat{T}$, $\hat{S}$, and $\hat{P}$, which exchange the time, spin, and coordinate of the paired particles as
\begin{align}
&\hat{T}\mathcal{F}_{ab}({\bm r}_1,{\bm r}_2;\tau) = \mathcal{F}_{ba}({\bm r}_1,{\bm r}_2;\tau), \\
&\hat{S}\mathcal{F}_{ab}({\bm r}_1,{\bm r}_2;\tau) = \mathcal{F}_{ab}({\bm r}_1,{\bm r}_2;-\tau), \\
&\hat{P}\mathcal{F}_{ab}({\bm r}_1,{\bm r}_2;\tau) = \mathcal{F}_{ab}({\bm r}_2,{\bm r}_1;\tau).
\end{align}
We notice that the order-two operators show $\hat{T}^2=1$, $\hat{S}^2=1$, and $\hat{P}^2=1$, and the Fermi-Dirac statistics requires $\hat{T}\hat{S}\hat{P}=-1$. In the case of single-band SCs/SFs, therefore, exit the $2\times 2\times 2 /2 = 4$ classes of the possible Cooper pair correlations in terms of the parity of the coordinate, time, and spin exchanges: Even-frequency spin-singlet even-parity (ESE, $\hat{T}=+1,\hat{S}=-1,\hat{P}=+1$), even-frequency spin-triplet odd-parity (ETO, $\hat{T}=+1,\hat{S}=+1,\hat{P}=-1$), odd-frequency spin-singlet odd-parity (OSO, $\hat{T}=-1,\hat{S}=-1,\hat{P}=-1$), and odd-frequency spin-triplet even-parity (OTE, $\hat{T}=-1,\hat{S}=+1,\hat{P}=+1$) pairings. 

In the case of spin-triplet SCs/SFs, such as $^3$He, only ETO pairing exists in the bulk. A time-reversal breaking perturbation can induce the mixing of spin-single pairing. Since the spin-singlet pairing must have an odd parity unless the translational symmetry is broken, OSO pairing is emergent in bulk ETO SCs/SFs and ESE pairing is prohibited by the symmetry. Once a translational symmetry is broken by the existence of an interface, surface, and vortices, the symmetry breaking induces OTE pairs in ETO SCs/SFs. All four pairings can emerge when both time-reversal symmetry and translational symmetry are broken. The concept of odd frequency pairing has succeeded in extracting the fundamental roles of ABSs in anomalous spin and charge transport, electromagnetic responses, proximity effects~\cite{tanakaJPSJ12,eschrigJLTP07}. In particular, odd-frequency pairing emergent in SCs/SFs is found to yield anomalous paramagnetic responses and negative superfluid density~\cite{asanoPRB14,higashitaniPRB14}. 

Using the quasiclassical approximation reliable for $T_{\rm c}/T_{\rm F} \ll 1$, Higashitani {\it et al.}~\cite{higashitaniPRB12} and Tsutsumi and Machida~\cite{tsutsumi:2012c} demonstrated that in $^3$He-B, the momentum resolved surface density of states, $\mathcal{N}(\hat{\bm k},z;E)$, is identical to the OTE pair amplitude as
\beq
\mathcal{N}(\hat{\bm k},z=z_{\rm surf};E)
\approx \frac{1}{\pi}
\left|{\rm Re}f^{\rm OF}_z(\hat{\bm k},z=z_{\rm surf};\omega _n \rightarrow -iE+0_+)\right|.
\label{eq:equality}
\eeq
for $|E|<\Delta _{\rm B}$. The OTE pairing amplitude, $f^{\rm OF}_z$, is defined by the quasiclassical approximation of the anomalous Green's function, $f_{ab}(\hat{\bm k},z;\omega _n)$, as $f^{\rm OF}_{\mu}={\rm tr}[-i\sigma _y \sigma _{\mu}(f(\omega _n)-f(\omega _n))]/4$. Hence, the surface ABS has multifaceted properties as odd-frequency pairing and Majorana fermions. The remarkable consequence of the identity between the odd-frequency pairing amplitude and surface ABSs in $^3$He-B is the anomalous enhancement of surface spin susceptibility $\chi _{\rm surf}$ around the topological quantum critical point $H^{\ast}$ in Fig.~\ref{fig:phase}(a). For time-reversal-invariant spin-triplet (ETO) SCs/SFs, in general, the surface spin susceptibility is composed of the contributions of odd-parity (ETO/OSO) pair amplitudes, $\chi^{\rm OP}_{\rm surf}$, and enen-parity (OTE/ESE) pair amplitudes, $\chi^{\rm EP}_{\rm surf}$, as $\chi _{\rm surf}=\chi _{\rm N}+\chi^{\rm OP}_{\rm surf} + \chi^{\rm EP}_{\rm surf}$. Mizushima~\cite{mizushimaPRB14} found that the emergent Cooper pair amplitudes and surface spin susceptibility are subject to the discrete symmetries that the superfluid/superconducting phases hold. In the $^3$He-B under a magnetic field, the surface spin susceptibility is expressed in terms of the Ising order $\hat{\ell}_z\in[-1,1]$ as 
\beq
\chi _{\rm surf} 
= \chi _{\rm N} + \sqrt{1-\hat{\ell}^2_{z}}\tilde{\chi}^{\rm OP}_{\rm surf}
+ \hat{\ell}_z \tilde{\chi}^{\rm EP}_{\rm surf},
\label{eq:chi_final}
\eeq
where $\tilde{\chi}$ is the spin susceptibility of the BW state with the simplest form of the oder parameter, $\hat{\bm n}=\hat{\bm z}$ and $\varphi =0$. The contribution of odd-parity pairings, $\chi^{\rm OP}_{\rm surf}$, is understandable with the orientation of the ${\bm d}$-vector relative to the applied field $\hat{\bm h}$: $\chi^{\rm OP}=0$ for $\hat{\bm h}\perp{\bm d}$ and $\chi^{\rm OP} \le 0$ for ${\bm d}\cdot\hat{\bm h}\neq 0$. Since the order parameter of the BW state has three ${\bm d}$-vectors, the odd parity contribution always suppresses the spin susceptibility $\chi ^{\rm OP}<0$. The even-parity contribution is estimated within the Ginzburg-Landau regime as $\chi^{\rm EP}_{\rm surf} = \frac{7\zeta(3)}{12(1+F^{\rm a}_0)}( \Delta _{\rm B}/\pi T)^2   > 0$~\cite{mizushimaPRB14,mizushimaJPCM15}, where $\zeta(3)$ is the Riemann zeta function and $F^{\rm a}_0 \approx -0.7$ denotes the Landau's Fermi liquid parameter. Hence, Eq.~\eqref{eq:chi_final} clearly shows that only the even-parity (OTE/ESE) pairs contribute to the surface spin susceptibility when $\hat{\ell}_z=0$ (the $P_3$ symmetric topological ${\rm B}_{\rm I}$ phase), while $\chi$ for $\hat{\ell}_z=1$ (the non-topological ${\rm B}_{\rm II}$ phase) is composed of only the odd-parity (ETO/OSO) Cooper pairs,
\beq
\chi = \left\{
\begin{array}{ll}
\displaystyle{\chi _{\rm N}+ \tilde{\chi}^{\rm OP} < \chi _{\rm N}} & \mbox{for $\hat{\ell}_z = 0$} \\
\\
\displaystyle{\chi _{\rm N}+ \tilde{\chi}^{\rm EP} > \chi _{\rm N}} & \mbox{for $\hat{\ell}_z = 1$} 
\end{array}
\right. .
\eeq
The symmetry consideration described above implies that although the OTE pair amplitudes always exist in the surface of $^3$He-B, they do not couple to the applied magnetic field when $\hat{\ell}_z=0$. This is consistent to the Majorana Ising spin of massless Majorana fermions realized in the $P_3$ symmetric topological ${\rm B}_{\rm I}$ phase. In the ${\rm B}_{\rm II}$ phase for $H_{\parallel}>H^{\ast}$, owing to $\hat{\ell}_z\neq 0$, the even-parity contribution $\chi^{\rm EP}$ abruptly increases and $\chi^{\rm OP}$ disappears. The paramagnetic response of the OTE pair amplitudes gives rise to the anomalous enhancement of the surface spin susceptibility that exceeds that of the normal state, $\chi _{\rm surf} > \chi _{\rm N}$. Since the $H_{\parallel}$-dependence of the surface susceptibility in $^3$He-B follows that of the Ising order $\hat{\ell}_z$, the measurement of the spin susceptibility in NMR experiments may detect the field-dependence of the mass acquisition and topological phase transition associated with the spontaneous symmetry breaking.

\section{Conclusions}
\label{sec:conclusions}

We have reviewed surface and vortex Andreev bound states in unconventional superconductors and superfluids and their multifaceted properties as Majorana fermions and odd-frequency pairing from the viewpoint of symmetry and topology. The unified description on the basis of the one-dimensional Dirac equation uncovers the fundamental roles of low-lying Andreev bound states on the thermodynamic stability of Fulde-Ferrell-Larkin-Ovchinnikov superconductors. In addition, the physical origin of the anomalous band structures of Andreev bound states in Pauli-limited two-band superconductors has been clarified in Sec.~2(a).  

The main part of the current review is devoted to giving an overview on the physics of Andreev bound states woven by the intertwining of topology and symmetry. The special focus has been placed on the superfluid $^3$He as a prototype of Weyl superconductors and symmetry-protected topological phases. In particular, it is recognized that the $^3$He-B confined in a slab geometry can serve as a unique platform to realize the {\it topological quantum criticality} as displayed in Fig.~\ref{fig:phase}(a), which may contain new quantum phenomena such as spontaneous-symmetry-breaking-driven mass acquisition of ABSs, Majorana Ising spins, and emergent supersymmetry. 

In $^3$He, the existence of gapless surface bound states was first pointed out by Buchholtz and Zwicknagl in 1981~\cite{buchholtz:1981}. They uncovered the novel quasiparticle states by numerically solving the quasiclassical self-consistent theory with or without surface roughness. After the pioneering work, the efforts to detect ABSs in $^3$He have been made by numerous experimental groups. The manifestations of surface Andreev bound states have been captured by several experimental groups through the heat capacity measurement~\cite{choi,bunkov}, anomalous attenuation of transverse sound waves~\cite{davisPRL08}, and transverse acoustic impedance~\cite{okuda,aokiPRL05, saitoh, wada, murakawaPRL09, murakawaJPSJ11}. Among them, Murakawa {\it et al.} succeeded in the measurement of the transverse acoustic impedance with a controllable surface condition, which serves as a high resolution spectroscopy of surface Andreev bound states. Indeed, they observed the transformation of the characteristic shape of surface density of states by changing the surface condition from the diffusive to specular limit. Hence, a pressing issue shared by extensive field of condensed matter physics is to capture a hallmark of Majorana nature of surface Andreev bound states such as the Majorana Ising spins, spontaneous-symmetry-breking-driven mass acquisition, and topological quantum criticality as discussed in Sec.~3. 

\enlargethispage{20pt}



\aucontribute{
Both authors contributed equally to this work.
}

\competing{The author declare that they have no competing interests.}

This work was supported by JSPS (Nos.~25800199, 25287085, and JP16K05448) and ``Topological Quantum Phenomena'' (No.~22103005) and ``Topological Materials Science'' (No.~15H05855) KAKENHI on innovation areas from MEXT. 

\ack{We gratefully thank M. Sato, Y. Tsutsumi, T. Kawakami, M. Ichioka, M. Takahashi, S. Higashitani, S. Fujimoto, J. A. Sauls, and Y. Tanaka for fruitful collaborations and helpful and stimulating discussions.}





\begin{thebibliography}{9}

\bibitem{kashiwayaRPP}
Kashiwaya S, Tanaka Y. 2000. Tunnelling effects on surface bound states in unconventional superconductors.
\textit{Rep. Prog. Phys.} \textit{63} 1641.

\bibitem{tanakaJPSJ12}
Tanaka Y, Sato M, Nagaosa N. 2012. Symmetry and Topology in Superconductors-Odd-Frequency Pairing and Edge States-.
{\it J. Phys. Soc. Jpn.} {\bf 81} 011013.

\bibitem{nagaiJPSJ08}
Nagai K, Nagato Y, Yamamoto M, Higashitani S. 2008. Surface Bound States in Superfluid $^3$He.
{\it J. Phys. Soc. Jpn.} {\bf 77} 111003.

\bibitem{ff}
Fulde P, Ferrell RA. 1964. Superconductivity in a Strong Spin-Exchange Field.
{\it Phys. Rev.} \textbf{135} A550.

\bibitem{lo}
Larkin AI, Ovchinnikov YN. 1965. Inhomogeneous state of superconductors.
{\it Sov. Phys. JETP} \textbf{20} 762.

\bibitem{matsudaJPSJ07}
Matsuda Y, Shimahara H. 2007.
Fulde-Ferrell-Larkin-Ovchinnikov State in Heavy Fermion Superconductors.
{\it J. Phys. Soc. Jpn.} {\bf 76} 051005.


\bibitem{aokiPRL05}
Aoki Y, Wada Y, Saitoh M, Nomura R, Okuda Y, Nagato Y, Yamamoto M, Higashitani S, Nagai K. 2005.
Observation of Surface Andreev Bound States of Superfluid $^{3}\mathrm{He}$ by Transverse Acoustic Impedance Measurements.
{\it Phys. Rev. Lett.} {\bf 95} 075301.

\bibitem{saitoh}
Saitoh M, Wada Y, Aoki Y, Murakawa S, Nomura R, Okuda Y. 2006.
Spectroscopic study of the surface density of states of superfluid $^{3}\mathrm{He}$ by transverse acoustic impedance measurements.
{\it Phys. Rev. B} {\bf 74} 220505.

\bibitem{wada}
Wada Y, Murakawa S, Tamura Y, Saitoh M, Aoki Y, Nomura R, Okuda, Y. 2008.
Broadening of the surface Andreev bound states band of superfluid $^{3}\text{H}\text{e-}B$ on a partially specular wall.
{\it Phys. Rev. B} {\bf 78} 214516.

\bibitem{murakawaPRL09}
Murakawa S, Tamura Y, Wada Y, Wasai M, Saitoh M, Aoki Y, Nomura R, Okuda Y, Nagato Y,  Yamamoto M, Higashitani S, Nagai K. 2009,
New Anomaly in the Transverse Acoustic Impedance of Superfluid $^{3}\mathrm{He}\mathrm{\text{-}}\mathbf{B}$ with a Wall Coated by Several Layers of $^{4}\mathrm{He}$.
{\it Phys. Rev. Lett.} {\bf 103} 155301.

\bibitem{murakawaJPSJ11}
Murakawa S, Wada Y, Tamura Y, Wasai M, Saitoh M, Aoki Y, Nomura R, Okuda Y, Nagato Y, Yamamoto M, Higashitani S, Nagai K. 2011.
Surface Majorana Cone of the Superfluid $^3$He-B Phase.
{\it J. Phys. Soc. Jpn.} {\bf 80} 013602.

\bibitem{mizushimaJPCM15}
Mizushima T, Tsutsumi Y, Sato M, Machida K. 2015. Symmetry protected topological superfluid ${}^3$He-B.
\textit{J. Phys.:Condens. Matter} \textbf{27} 113203.

\bibitem{mizushimaJPSJ15}
Mizushima T, Tsutsumi Y, Kawakami T, Sato M, Ichioka M, Machida K. 2016. 
Symmetry Protected Topological Superfluids and Superconductors-From the Basics to $^3$He-.
{\it J. Phys. Soc. Jpn.} {\bf 85} 022001.

\bibitem{okuda}
Okuda Y, Nomura R. 2012.
Surface Andreev bound states of superfluid $^3$He and Majorana fermions.
{\it J. Phys.:Condens. Matter} {\bf 24} 343201.

\bibitem{volovik} Volovik GE. 2003 \textit{The Universe in a Helium Droplet}.
Oxford, UK: Oxford University Press. 

\bibitem{covington}
Covington M, Aprili M, Paraoanu E, Greene LH, Xu F, Zhu J, and Mirkin CA. 1997.
Observation of Surface-Induced Broken Time-Reversal Symmetry in ${\mathrm{YBa}}_{2}{\mathrm{Cu}}_{3}{O}_{7}$ Tunnel Junctions.
{\it Phys. Rev. Lett.} {\bf 79} 277.

\bibitem{kashiwaya}
Kashiwaya S, Tanaka Y, Koyanagi M, Takashima H, Kajimura K. 1995.
Origin of zero-bias conductance peaks in high-${\mathit{T}}_{\mathit{c}}$ superconductors.
{\it Phys. Rev. B} {\bf 51} 1350.

\bibitem{leseur}
Lesueur J, Greene LH, Feldmann WL, Inam A. 1992.
Zero bias anomalies in ${\mathrm{YBa}}_{2}{\mathrm{Cu}}_{3}{O}_{7}$ tunnel junctions.
{\it Physica C: Superconductivity} {\bf 191} 325.


\bibitem{hess}
Hess HF, Robinson RB, Dynes RC, Valles JM, Waszczak JV. 1989.
Scanning-Tunneling-Microscope Observation of the Abrikosov Flux Lattice and the Density of States near and inside a Fluxoid.
{\it Phys. Rev. Lett.} {\bf 62} 214.




\bibitem{wilczek}
Wilczek F. 2009. Majorana returns. {\it Nat. Phys.} {\bf 5} 614.

\bibitem{qiRMP11}
Qi XL, Zhang SC. 2011. Topological insulators and superconductors.
{\it Rev. Mod. Phys.} {\bf 83} 1057.

\bibitem{ueno}
Ueno Y, Yamakage A, Tanaka Y, Sato M. 2013.
Symmetry-Protected Majorana Fermions in Topological Crystalline Superconductors: Theory and Application to Sr$_2$RuO$_4$.
{\it Phys. Rev. Lett.} {\bf 111} 087002.

\bibitem{sato14}
Sato M, Yamakage A, Mizushima T. 2014.
Mirror {M}ajorana zero modes in spinful superconductors/superfluids Non-{A}belian anyons in integer quantum vortices.
{\it Physica E} {\bf 55} 20.

\bibitem{liuPRX14}
Liu XJ, Wong CLM, Law KT. 2014.
Non-Abelian Majorana Doublets in Time-Reversal-Invariant Topological Superconductors.
{\it Phys. Rev. X} {\bf 4} 021018.

\bibitem{chungPRL09}
Chung SB, Zhang SC. 2009. 
Detecting the Majorana Fermion Surface State of $^{3}\mathrm{He}\mathrm{\text{-}}B$ through Spin Relaxation.
{\it Phys. Rev. Lett.} {\bf 103} 235301.

\bibitem{nagatoJPSJ09}
Nagato Y, Higashitani S, Nagai K. 2009.
Strong Anisotropy in Spin Susceptibility of Superfluid $^3$He-B Film Caused by Surface Bound States.
{\it J. Phys. Soc. Jpn.} {\bf 78} 123603.

\bibitem{shindouPRB10}
Shindou R, Furusaki A, Nagaosa N. 2010. Quantum impurity spin in Majorana edge fermions.
{\it Phys. Rev. B} {\bf 82} 180505.

\bibitem{mizushimaPRL12}
Mizushima T, Sato M, Machida K. 2012. 
Symmetry Protected Topological Order and Spin Susceptibility in Superfluid $^{3}\mathrm{He}\mathrm{\text{-}}B$.
{\it Phys. Rev. Lett.} {\bf 109} 165301.

\bibitem{mizushimaPRB12} Mizushima T. 2012.  Superfluid ${}^{3}$He in a restricted geometry with a perpendicular magnetic field. \textit{Phys. Rev. B} \textbf{86} 094518.

\bibitem{shiozakiPRB14}
Shiozaki K, Sato M. 2014. Topology of crystalline insulators and superconductors.
{\it Phys. Rev. B} {\bf 90} 165114.

\bibitem{eschrigJLTP07}
Eschrig M, L\"{o}fwander T, Champel T, Cuevas JC, Kopu J, Sch\"{o}n G. 2007
Symmetries of Pairing Correlations in Superconductor-Ferromagnet Nanostructures.
{\it J. Low Temp. Phys.} {\bf 147} 457.

\bibitem{tanakaPRL07}
Tanaka Y, Golubov AA. 2007.
Theory of the Proximity Effect in Junctions with Unconventional Superconductors.
{\it Phys. Rev. Lett.} {\bf 98} 037003.

\bibitem{tanakaPRL07-2}
Tanaka Y, Golubov AA, Kashiwaya S, Ueda M. 2007.
Anomalous Josephson Effect between Even- and Odd-Frequency Superconductors.
{\it Phys. Rev. Lett.} {\bf 99} 037005.

\bibitem{eschrigNP08}
Eschrig M, L\"{o}fwander T. 2008. 
Triplet supercurrents in clean and disordered half-metallic ferromagnets.
{\it Nat. Phys.} {\bf 4} 138-143.

\bibitem{fominov}
Fominov YV, Tanaka Y, Asano Y, Eschrig M. 2015.
Odd-frequency superconducting states with different types of Meissner response: Problem of coexistence.
{\it Phys. Rev. B} {\bf 91} 144514.

\bibitem{matsumoto}
Matsumoto M, Koga M, Kusunose H. 2013.
Emergent Odd-Frequency Superconducting Order Parameter near Boundaries in Unconventional Superconductors.
{\it J. Phys. Soc. Jpn.} {\bf 82} 034708.

\bibitem{shigeta}
Shigeta K, Onari S, Tanaka Y. 2013.
Possible Odd-Frequency Pairing in Quasi-One-Dimensional Organic Superconductors (TMTSF)$_2X$.
{\it J. Phys. Soc. Jpn.} {\bf 82} 104702.

\bibitem{hoshino}
Hoshino S, Kuramoto Y. 2014.
Superconductivity of Composite Particles in a Two-Channel Kondo Lattice.
{\it Phys. Rev. Lett.} {\bf 112} 167204.

\bibitem{mizushimaPRB14} Mizushima T. 2014. Odd-frequency pairing and Ising spin susceptibility in time-reversal-invariant superfluids and superconductors. \textit{Phys. Rev. B} \textbf{90} 184506.


\bibitem{gross}
Gross DJ. 1974. Dynamical symmetry breaking in asymptotically free field theories. 
\textit{Phys. Rev. D} \textbf{10} 3235.

\bibitem{nambuPR1961}
Nambu Y, Jona-Lasinio G. 1961. Dynamical Model of Elementary Particles Based on an Analogy with Superconductivity. I.
\textit{Phys. Rev.} \textbf{122} 345.

\bibitem{jackiw}
Jackiw R, Rebbi C. 1976. Solitons with fermion number $1/2$. \textit{Phys. Rev. D} \textbf{13} 3398.

\bibitem{akns}
Correa F, Dunne GV, Plyushchay MS. 2009. The Bogoliubov-de Gennes system, the AKNS hierarchy, and nonlinear quantum mechanical supersymmetry.
\textit{Ann. Phys.} \textbf{324} 2522.

\bibitem{hu}
Hu CR. 1975. Rigorous study of the gap equation for an inhomogeneous superconducting state near ${T}_{c}$.
\textit{Phys. Rev. B} \textbf{12} 3635.

\bibitem{bar-sagi}
Bar-Sagi J, Kuper CG. 1972. Self-Consistent Pair Potential in an Inhomogeneous Superconductor. 
\textit{Phys. Rev. Lett.} \textbf{28} 1556.

\bibitem{shei}
Shei SS. 1976. Semiclassical bound states in a model with chiral symmetry. \textit{Phys. Rev. D} \textbf{14} 535.

\bibitem{campbell}
Campbell DK, Bishop AR. 1981. Solitons in polyacetylene and relativistic-field-theory models. 
\textit{Phys. Rev. B} \textbf{24} 4859.

\bibitem{okuno}
Okuno S, Onodera Y. 1983. Coexistence of a Soliton and a Polaron in Trans-Polyacetylene. 
\textit{J. Phys. Soc. Jpn.} \textbf{52} 3495.

\bibitem{feinberg2003}
Feinberg J. 2003. Marginally stable topologically non-trivial solitons in the Gross-Neveu model.
\textit{Phys. Lett. B} \textbf{569} 204.

\bibitem{feinberg2004}
Feinberg J. 2004. All about the static fermion bags in the Gross-Neveu model.
\textit{Ann. Phys.} \textbf{309} 166.

\bibitem{basar1}
Ba\ifmmode \mbox{\c{s}}\else \c{s}\fi{}ar G, Dunne GV. 2008. Self-Consistent Crystalline Condensate in Chiral Gross-Neveu and Bogoliubov-de Gennes Systems. \textit{Phys. Rev. Lett.} \textbf{100} 200404.

\bibitem{basar2}
Ba\ifmmode \mbox{\c{s}}\else \c{s}\fi{}ar G, Dunne GV. 2008. Twisted kink crystal in the chiral Gross-Neveu model.
\textit{Phys. Rev. D} \textbf{78} 065022.

\bibitem{basar3}
Ba\ifmmode \mbox{\c{s}}\else \c{s}\fi{}ar G, Dunne GV, Thies M. 2009. Inhomogeneous condensates in the thermodynamics of the chiral ${\mathrm{NJL}}_{2}$ model.
\textit{Phys. Rev. D} \textbf{79} 105012.

\bibitem{takahashi}
Takahashi DA, Nitta M. 2013. Self-Consistent Multiple Complex-Kink Solutions in Bogoliubov\char21{}de~Gennes and Chiral Gross-Neveu Systems.
\textit{Phys. Rev. Lett.} \textbf{110} 131601.

\bibitem{brazovskii}
Brazovskii SA, Gordynin SA, Kirova NN. 1980. An exact solution of the Peierls model with an arbitrary number of electrons in the unit cell.
\textit{Pis'ma Zh. Eksp. Teor. Fiz.} \textbf{31} 486 [{\it JETP Lett.} {\bf 31} 456].

\bibitem{mertsching}
Mertsching J, Fischbeck, HJ. 1981. The Incommensurate Peierls Phase of the Quasi-One-Dimensional Fr\"{o}hlich Model with a Nearly Half-Filled Band.
\textit{Phys. Status Solidi B} \textbf{103} 783.

\bibitem{horovitz}
Horovitz B. 1981. Soliton Lattice in Polyacetylene, Spin-Peierls Systems, and Two-Dimensional Sine-Gordon Systems
\textit{Phys. Rev. Lett.} \textbf{46} 742.

\bibitem{machidaPRB1984-2}
Machida K, Fujita M. 1984. Soliton lattice structure of incommensurate spin-density waves: Application to Cr and Cr-rich Cr-Mn and Cr-V alloys.
\textit{Phys. Rev. B} \textbf{30} 5284.

\bibitem{fujitaJPSJ1984}
Fujita M, Machida K. 1984. Spin-Peierls Transitions in Magnetic Fields-Thermodynamic Properties of a Soliton Lattice State-.
\textit{J. Phys. Soc. Jpn.} \textbf{53} 4395.

\bibitem{machida1989}
Machida K. 1989. Magnetism in La$_2$CuO$_4$ based compounds.
\textit{Physica (Amsterdam)} {\bf 158C} 192.

\bibitem{machidaPRB1984}
Machida K, Nakanishi H. 1984. Superconductivity under a ferromagnetic molecular field.
\textit{Phys. Rev. B} {\bf 30} 122. 

\bibitem{yoshii}
Yoshii R, Tsuchiya S, Marmorini G, Nitta M. 2011. Spin imbalance effect on the Larkin-Ovchinnikov-Fulde-Ferrel state.
\textit{Phys. Rev. B} \textbf{84} 024503.

\bibitem{takayama}
Takayama H, Lin-Liu YR, Maki K. 1980. Continuum model for solitons in polyacetylene.
\textit{Phys. Rev. B} \textbf{21} 2388.

\bibitem{mizushimaPRL05-1}
Mizushima T, Machida K, Ichioka M. 2005. Direct Imaging of Spatially Modulated Superfluid Phases in Atomic Fermion Systems.
{\it Phys. Rev. Lett.} {\bf 94} 060404.

\bibitem{yokoyamaJPSJ10}
Yokoyama T, Ichioka M, Tanaka Y. 2010. 
Theory of Pairing Symmetry in Fulde-Ferrell-Larkin-Ovchinnikov Vortex State and Vortex Lattice. 
{\it J. Phys. Soc. Jpn.} {\bf 79} 034702.

\bibitem{higashitaniPRB14}
Higashitani S. 2014. 
Odd-frequency pairing effect on the superfluid density and the Pauli spin susceptibility in spatially nonuniform spin-singlet superconductors.
{\it Phys. Rev. B} {\bf 89} 184505.

\bibitem{wang}
Wang Q, Chen HY, Hu CR, Ting CS. 2006.
Local Tunneling Spectroscopy as a Signature of the Fulde-Ferrell-Larkin-Ovchinnikov State in $s$- and $d$-Wave Superconductors.
{\it Phys. Rev. Lett.} {\bf 96} 117006.

\bibitem{anton}
Vorontsov AB, Sauls JA, Graf MJ. 2005. 
Phase diagram and spectroscopy of Fulde-Ferrell-Larkin-Ovchinnikov states of two-dimensional $d$-wave superconductors.
{\it Phys. Rev. B} {\bf 72} 184501.

\bibitem{ichiokaPRB07-2}
Ichioka M, Adachi H, Mizushima T, Machida K. 2007.
Vortex state in a Fulde-Ferrell-Larkin-Ovchinnikov superconductor based on quasiclassical theory.
{\it Phys. Rev. B} {\bf 76} 014503.

\bibitem{suzukiJPSJ11}
Suzuki KM, Tsutsumi Y, Nakai N, Ichioka M, Machida K. 2011.
Field Evolution of the Fulde–Ferrell–Larkin–Ovchinnikov State in a Superconductor with Strong Pauli Effects.
{\it J. Phys. Soc. Jpn.} {\bf 80} 123706.

\bibitem{mizushimaJPSJ14}
Mizushima T, Takahashi M, Machida K. 2014. Fulde-Ferrell-Larkin-Ovchinnikov States in Two-Band Superconductors.
\textit{J. Phys. Soc. Jpn.} \textbf{83} 023703. 

\bibitem{takahashiPRB14}
Takahashi M, Mizushima T, Machida K. 2014. Multiband effects on Fulde-Ferrell-Larkin-Ovchinnikov states of Pauli-limited superconductors. \textit{Phys. Rev. B} \textbf{89} 064505.

\bibitem{silaev:2012}
Silaev MA, Volovik GE. 2012. Topological Fermi arcs in superfluid ${}^{3}$He.
{\it Phys. Rev. B} {\bf 86} 214511.

\bibitem{read}
Read N, Green D. 2000.
Paired states of fermions in two dimensions with breaking of parity and time-reversal symmetries and the fractional quantum Hall effect.
{\it Phys. Rev. B} {\bf 61} 10267-10297.

\bibitem{mizushimaPRL08}
Mizushima T, Ichioka M, Machida K. 2008. Role of the Majorana Fermion and the Edge Mode in Chiral Superfluidity near a $p$-Wave Feshbach Resonance.
{\it Phys. Rev. Lett.} {\bf 101} 150409.

\bibitem{balentsPRB12}
Meng T, Balents L. 2012. Weyl superconductors. {\it Phys. Rev. B} {\bf 86} 054504.

\bibitem{sauPRB12}
Sau JD, Tewari S. 2012. Topologically protected surface Majorana arcs and bulk Weyl fermions in ferromagnetic superconductors.
{\it Phys. Rev. B} {\bf 86} 104509.

\bibitem{dasPRB13}
Das T. 2013. Weyl semimetal and superconductor designed in an orbital-selective superlattice.
{\it Phys. Rev. B} {\bf 88} 035444.

\bibitem{yangPRL14}
Yang SA, Pan H, Zhang F. 2014. Dirac and Weyl Superconductors in Three Dimensions.
{\it Phys. Rev. Lett.} {\bf 113} 046401.

\bibitem{silaevJETP14}
Silaev MA, Volovik GE. 2014. Andreev-Majorana bound states in superfluids.
{\it J. Exp. Theor. Phys.} {\bf 119} 1042. 


\bibitem{abm1}
Anderson PW, Morel P. 1961.
Generalized Bardeen-Cooper-Schrieffer States and the Proposed Low-Temperature Phase of Liquid ${\mathrm{He}}^{3}$.
{\it Phys. Rev.} {\bf 123} 1911--1934.

\bibitem{abm2}
Anderson PW, Brinkman WF. 1973. 
Anisotropic Superfluidity in $^{3}\mathrm{He}$: A Possible Interpretation of Its Stability as a Spin-Fluctuation Effect.
{\it Phys. Rev. Lett.} {\bf 30} 1108-1111.

\bibitem{tsutsumi:2010b}
Tsutsumi Y, Mizushima T, Ichioka M, Machida K. 2010.
Majorana Edge Modes of Superfluid $^3$He A-Phase in a Slab.
{\it J. Phys. Soc. Jpn.} {\bf 79} 113601.

\bibitem{tsutsumi:2011b}
Tsutsumi Y, Ichioka M, Machida K. 2011. 
Majorana surface states of superfluid $^{3}He$ $A$ and $B$ phases in a slab.
{\it Phys. Rev. B} {\bf 83} 094510.

\bibitem{ikegami1}
Ikegami H, Tsutsumi Y, Kono K. 2013. 
Chiral Symmetry Breaking in Superfluid $^3$He-A.
{\it Science} {\bf 341} 59-62.

\bibitem{ikegami2}
Ikegami H, Tsutsumi Y, Kono K. 2015.
Observation of Intrinsic Magnus Force and Direct Detection of Chirality in Superfluid $^3$He-A.
{\it J. Phys. Soc. Jpn.} {\bf 84} 044602.

\bibitem{salmelin1}
Salmelin RH, Salomaa MM, Mineev VP. 1989.
Internal Magnus effects in superfluid $^3$He-A
{\it Phys. Rev. Lett.} {\b 63} 868.

\bibitem{salmelin2}
Salmelin RH, Salomaa MM. 1990.
Resonant quasiparticle-ion scattering in anisotropic superfluid $^3$He.
{\it Phys. Rev. B} {\bf 41} 4142.

\bibitem{yamashita}
Yamashita T, Shimoyama Y, Haga Y, Matsuda TD, Yamamoto E, {\=O}nuki Y, Sumiyoshi H, Fujimoto S, Levchenko A, Shibauchi T, Matsuda Y. 2015. 
Colossal thermomagnetic response in the exotic superconductor ${\mathrm{URu}}_{2}{\mathrm{Si}}_{2}$.
{\it Nat. Phys.} {\bf 11}, 17.

\bibitem{schemmPRB15}
Schemm ER, Baumbach RE, Tobash PH, Ronning F, Bauer ED, Kapitulnik A. 2015.
Evidence for broken time-reversal symmetry in the superconducting phase of ${\mathrm{URu}}_{2}{\mathrm{Si}}_{2}$.
{\it Phys. Rev. B} {\bf 91} 140506.

\bibitem{mineev}
Mineev VP. 2002.
Superconducting states in ferromagnetic metals.
{\it Phys. Rev. B} {\bf 66} 134504.

\bibitem{shimizu}
Shimizu Y, Kittaka S, Nakamura T, Sakakibara T, Aoki D, Homma Y, Nakamura A, Kachida K. 2017.
Quasiparticle excitations and evidence for superconducting double transitions in monocrystalline U$_{0.97}$Th$_{0.03}$Be$_{13}$. 
{\it Phys. Rev. B} {\bf 96} 100505.

\bibitem{mizushimaPRB18}
Mizushima T, Nitta M. 2018
Topology and symmetry of surface Majorana arcs in cyclic superconductors.
{\it Phys. Rev. B} {\bf 97} 024506. 

\bibitem{machidaJPSJ18}
Machida K. 2018. 
Spin Triplet Nematic Pairing Symmetry and Superconducting Double Transition in U$_{1-x}$Th$_x$Be$_{13}$.
{\it J. Phys. Soc. Jpn.} {\bf 87} 033703.

\bibitem{joynt}
Joynt R, Taillefer L. 2002.
The superconducting phases of ${\mathrm{UPt}}_{3}$.
{\it Rev. Mod. Phys.} {\bf 74} 235-294.

\bibitem{schemm14}
Schemm ER, Gannon WJ, Wishne CM, Halperin WP, Kapitulnik A. 2014.
Observation of broken time-reversal symmetry in the heavy-fermion superconductor UPt$_3$.
{\it Science} 345 190.

\bibitem{sumiyoshi}
Sumiyoshi H, Fujimoto S. 2014.
Giant Nernst and Hall effects due to chiral superconducting fluctuations.
{\it Phys. Rev. B} {\bf 90} 184518.

\bibitem{sauls94}
Sauls JA. 1994. The Order Parameter for the Superconducting Phases of UPt$_3$.
{\it Adv. Phys.} {\bf 43} 113.

\bibitem{tsutsumiJPSJ12-2}
Tsutsumi Y, Machida K, Ohmi T, Ozaki M. 2012.
A Spin Triplet Superconductor UPt$_3$. 
{\it J. Phys. Soc. Jpn.} {\bf 81} 074717.

\bibitem{machidaPRL12}
Machida Y, Itoh A, So Y, Izawa K, Haga Y, Yamamoto E, Kimura N, \={O}nuki Y, Tsutsumi Y, Machida K. 2012.
Twofold Spontaneous Symmetry Breaking in the Heavy-Fermion Superconductor ${\mathrm{UPt}}_{3}$.
{\it Phys. Rev. Lett.} {\bf 108} 157002.

\bibitem{izawa}
Izawa K, Machida Y, Itoh A, So Y, Ota K, Haga Y, Yamamoto E, Kimura N, {\=O}nuki Y, Tsutsumi Y, Machida K.
2014. Pairing Symmetry of UPt$_3$ Probed by Thermal Transport Tensors.
{\it J. Phys. Soc. Jpn.} {\bf 83} 061013.

\bibitem{goswami14}
Goswami P, Nevidomskyy AH. 2015.
Double Berry monopoles and topological surface states in the superconducting B-phase of UPt$_3$.
{\it Phys. Rev. B} {\bf 92} 214504.

\bibitem{tsutsumiJPSJ13}
Tsutsumi Y, Ishikawa M, Kawakami T, Mizushima T, Sato M, Ichioka M, Machida K. 2013. 
UPt$_3$ as a Topological Crystalline Superconductor.
{\it J. Phys. Soc. Jpn.} {\bf 82} 113707.

\bibitem{stone}
Stone M, Roy R. 2004. Edge modes, edge currents, and gauge invariance in ${p}_{x}{+ip}_{y}$ superfluids and superconductors.
{\it Phys. Rev. B} {\bf 69} 184511.

\bibitem{mizushimaPRA10}
Mizushima T, Machida K. 2010. Vortex structures and zero-energy states in the BCS-to-BEC evolution of $p$-wave resonant Fermi gases.
{\it Phys. Rev. A} {\bf 81} 053605.

\bibitem{sauls:2011}
Sauls JA. 2011. Surface states, edge currents, and the angular momentum of chiral $p$-wave superfluids.
{\it Phys. Rev. B} {\bf 84} 214509.

\bibitem{tsutsumi:2012}
Tsutsumi Y, Machida K. 2012. 
Edge mass current and the role of Majorana fermions in $A$-phase superfluid ${}^{3}$He
{\it Phys. Rev. B} {\bf 85} 100506(R).

\bibitem{tsutsumiJPSJ12}
Tsutsumi Y, Machida K. 2012. Edge Current due to Majorana Fermions in Superfluid $^{3}$He A- and B-Phases. 
{\it J. Phys. Soc. Jpn.} {\bf 81} 074607.

\bibitem{kopnin}
Kopnin N. 2001. {\it Theory of Nonequilibrium Superconductivity}.
Oxford, UK: Oxford University Press.

\bibitem{tewariPRL07}
Tewari S, Das Sarma S, Lee DH. 2007. Index Theorem for the Zero Modes of Majorana Fermion Vortices in Chiral $p$-Wave Superconductors.
{\it Phys. Rev. Lett.} {\bf 99} 037001.

\bibitem{satoPRB09-2}
Sato M, Fujimoto S. 2009.
Topological phases of noncentrosymmetric superconductors: Edge states, Majorana fermions, and non-Abelian statistics.
{\it Phys. Rev. B} {\bf 79} 094504.

\bibitem{cdgm}
Caroli C, De Gennes PG, Matricon J. 1964.
Bound Fermion states on a vortex line in a type \{II\} superconductor.
{\it Phys. Lett.} {\bf 9} 307.

\bibitem{kaneko}
Kaneko S, Matsuba K, Hafiz M, Yamasaki K, Kakizaki E, Nishida N, Takeya H, Hirata K, Kawakami T, Mizushima T, Machida K.
2012. Quantum Limiting Behaviors of a Vortex Core in an Anisotropic Gap Superconductor.
{\it J. Phys. Soc. Jpn.} {\bf 81} 063701.

\bibitem{hayashiPRL98}
Hayashi N, Isoshima T, Ichioka M, Machida K. 1998.
Low-Lying Quasiparticle Excitations around a Vortex Core in Quantum Limit.
{\it Phys. Rev. Lett.} {\bf 80} 2921.

\bibitem{hayashiJPSJ98}
Hayashi N, Ichioka M, Machida K. 1998. 
Relation between Vortex Core Charge and Vortex Bound States.
{\it J. Phys. Soc. Jpn.} {\bf 67} 3368.
 
\bibitem{mit}
Zwierlein MW, Abo-Shaeer JR, Schirotzek A, Schunck CH, Ketterle W. 2005. 
Vortices and Superfluidity in a Strongly Interacting Fermi Gas.
{\it Nature (London)} {\bf 435} 1047.

\bibitem{mizushimaPRL05}
Mizushima T, Ichioka M, Machida K. 2005. Topological Structure of a Vortex in the Fulde-Ferrell-Larkin-Ovchinnikov State. 
{\it Phys. Rev. Lett.} {\bf 95} 117003.


\bibitem{mizushimaPRA10-2}
Mizushima T, Machida K. 2010. 
Splitting and oscillation of Majorana zero modes in the $p$-wave BCS-BEC evolution with plural vortices.
{\it Phys. Rev. A} {\bf 82} 023624.

\bibitem{chengPRL09}
Cheng M, Lutchyn RM, Galitski V, Das Sarma S. 2009. 
Splitting of Majorana-Fermion Modes due to Intervortex Tunneling in a ${p}_{x}+i{p}_{y}$ Superconductor.
{\it Phys. Rev. Lett.} {\bf 103} 107001.

\bibitem{gurarie}
Gurarie V, Radzihovsky L. 2007. Zero modes of two-dimensional chiral $p$-wave superconductors.
{\it Phys. Rev. B} {\bf 75} 212509.

\bibitem{takahashiPRL06}
Takahashi M, Mizushima T, Machida K. 2006. Vortex-Core Structure in Neutral Fermion Superfluids with Population Imbalance. 
{\it Phys. Rev. Lett.} {\bf 97} 180407.

\bibitem{ichiokaPRB07}
Ichioka M, Machida K. 2007. Vortex states in superconductors with strong Pauli-paramagnetic effect.
{\it Phys. Rev. B} {\bf 76} 064502.


\bibitem{115}
Ikeda S, Shishido H, Nakashima M, Settai R, Aoki D, Haga Y, Harima H, Aoki Y, Namiki T, Sato H, {\=O}nuki Y. 
2001. Unconventional Superconductivity in CeCoIn$_5$ Studied by the Specific Heat and Magnetization Measurements.
{\it J. Phys. Soc. Jpn.} {\bf 77} 2248.

\bibitem{ube13}
Ramirez AP, Varma CM, Fisk Z, Smith JL. 1999. 
Fermi-liquid renormalization in the superconducting state of UBe$_{13}$.
{\it Philos. Mag. B} {\bf 79} 111.

\bibitem{shimizu}
Shimizu Y, Kittaka S, Sakakibara T, Haga Y, Yamamoto E, Amitsuka H, Tsutsumi Y, Machida K. 2015.
Field-Orientation Dependence of Low-Energy Quasiparticle Excitations in the Heavy-Electron Superconductor UBe$_{13}$.
{\it Phys. Rev. Lett.} {\bf 114} 147002.

\bibitem{kittaka}
Kittaka S, Aoki Y, Shimura Y, Sakakibara T, Seiro S, Geibel C, Steglich F, Ikeda H, Machida K.
2014. Multiband Superconductivity with Unexpected Deficiency of Nodal Quasiparticles in CeCu$_2$Si$_2$.
{\it Phys. Rev. Lett.} {112} 067002.

\bibitem{deguchi}
Deguchi K, Mao ZQ, Maeno Y.
2004. Determination of the Superconducting Gap Structure in All Bands of the Spin-Triplet Superconductor.
{\it J. Phys. Soc. Jpn.} {\bf 73} 1313.

\bibitem{suzukiPRA08}
Suzuki KM, Mizushima T, Ichioka M, Machida K. 2008. Magnetization profile and core-level spectroscopy in a multiply quantized vortex of imbalanced Fermi superfluids.
{\it Phys. Rev. A} {\bf 77} 063617.

\bibitem{eschrigPRB99}
Eschrig M, Sauls JA, Rainer D. 1999.
Electromagnetic response of a vortex in layered superconductors.
{\it Phys. Rev. B} {\bf 60} 10447.

\bibitem{saulsNJP09}
Sauls JA, Eschrig M. 2009.
Vortices in chiral, spin-triplet superconductors and superfluids.
{\it New J. Phys.} {\bf 11} 075008.

\bibitem{teoPRB10}
Teo JCY, Kane CL. 2010. 
Topological defects and gapless modes in insulators and superconductors.
{\it Phys. Rev. B} {\bf 82} 115120.

\bibitem{weinberg}
Weinberg EJ. 1981. 
Index calculations for the fermion-vortex system.
{\it Phys. Rev. D} {\bf 24} 2669-2673.

\bibitem{rossi}
Jackiw R, Rossi P. 1981. 
Zero modes of the vortex-fermion system.
{\it Nucl. Phys. B} {\bf 190} 681.

\bibitem{chengPRB10}
Cheng M, Lutchyn RM, Galitski V, Das Sarma, S. 2010.
Tunneling of anyonic Majorana excitations in topological superconductors.
{\it Phys. Rev. B} {\bf 82} 094504.

\bibitem{chiuPRB15}
Chiu CK, Pikulin DI, Franz M. 2015. 
Strongly interacting Majorana fermions.
{\it Phys. Rev. B} {\bf 91} 165402.

\bibitem{satoPL03}
Sato M. 2003.
Non-Abelian statistics of axion strings.
{\it Phys. Lett. B} {\bf 575} 126-130.

\bibitem{satoPRL09}
Sato M, Takahashi Y, Fujimoto S. 2009. 
Non-Abelian Topological Order in $s$-Wave Superfluids of Ultracold Fermionic Atoms.
{\it Phys. Rev. Lett.} {\bf 103} 020401.

\bibitem{satoPRB10-2}
Sato M, Takahashi Y, Fujimoto S. 2010. 
Non-Abelian topological orders and Majorana fermions in spin-singlet superconductors.
{\it Phys. Rev. B} {\bf 82} 134521. 

\bibitem{semenoff}
Semenoff GW, Sodano P. 2006. Stretching the Electron as Far as it Will Go. 
{\it Electron. J. Theor. Phys.} {\bf 10} 157-190.

\bibitem{ivanovPRL01}
Ivanov DA. 2001. Non-Abelian Statistics of Half-Quantum Vortices in $p$-Wave Superconductors.
{\it Phys. Rev. Lett.} {\bf 86} 268.

\bibitem{nayakRMP08}
Nayak C, Simon SH, Stern A, Freedman M, Das Sarma S. 2008.
Non-Abelian anyons and topological quantum computation.
{\it Rev. Mod. Phys.} {\bf 80} 1083.

\bibitem{chungPRL07}
Chung SB, Bluhm H, Kim EA. 2007.
Stability of Half-Quantum Vortices in ${p}_{x}+i{p}_{y}$ Superconductors.
{\it Phys. Rev. Lett.} {\bf 99} 197002.

\bibitem{kawakamiPRB09}
Kawakami T, Tsutsumi Y, Machida K. 2009.
Stability of a half-quantum vortex in rotating superfluid $^{3}\text{He}\text{-}A$ between parallel plates.
{\it Phys. Rev. B} {\bf 79} 092506.

\bibitem{vakaryukPRL09}
Vakaryuk V, Leggett AJ. 2009.
Spin Polarization of Half-Quantum Vortex in Systems with Equal Spin Pairing.
{\it Phys. Rev. Lett.} {\bf 103} 057003.

\bibitem{kawakamiJPSJ10}
Kawakami T, Tsutsumi Y, Machida K. 2010. 
Singular and Half-Quantum Vortices and Associated Majorana Particles in Superfluid $^3$He-A between Parallel Plates.
{\it J. Phys. Soc. Jpn.} {\bf 79} 044607.

\bibitem{kawakamiJPSJ11}
Kawakami T, Mizushima T, Machida K. 2011. 
Zero Energy Modes and Statistics of Vortices in Spinful Chiral $p$-Wave Superfluids.
{\it J. Phys. Soc. Jpn.} {\bf 80} 044603.

\bibitem{kondoJPSJ12}
Kondo K, Ohmi T, Nakahara M, Kawakami T, Tsutsumi Y, Machida K. 2012. 
Half-Quantum Vortices in Thin Film of Superfluid $^3$He.
{\it J. Phys. Soc. Jpn.} {\bf 81} 4603.


\bibitem{nakaharaPRB14}
Nakahara M, Ohmi T. 2014. Multiple half-quantum vortices in rotating superfluid $^3$He.
{\it Phys. Rev. B} {\bf 89} 104515.

\bibitem{yamashitaPRL08}
Yamashita M, Izumina K, Matsubara A, Sasaki Y, Ishikawa O, Takagi T, Kubota M, Mizusaki T. 2008. 
Spin Wave and Vortex Excitations of Superfluid $^{3}\mathrm{He}\mathrm{\text{-}}A$ in Parallel-Plate Geometry.
{\it Phys. Rev. Lett.} {\bf 101} 025302.

\bibitem{fangPRL14}
Fang C, Gilbert MJ, Bernevig BA. 2014.
New Class of Topological Superconductors Protected by Magnetic Group Symmetries.
{\it Phys. Rev. Lett.} {\bf 112} 106401.

\bibitem{liuprx14}
Liu XJ, Wong CLM, Law KT. 2014.
Non-Abelian Majorana Doublets in Time-Reversal-Invariant Topological Superconductors.
{\it Phys. Rev. X} {\bf 4} 021018.

\bibitem{wuPRB13}
Wu H, Sauls JA. 2013. 
Majorana excitations, spin and mass currents on the surface of topological superfluid ${}^{3}$He-B.
{\it Phys. Rev. B} {\bf 88} 184506.

\bibitem{schnyderPRB08}
Schnyder AP, Ryu S, Furusaki A, Ludwig AWW. 2008. Classification of topological insulators and superconductors in three spatial dimensions.
{\it Phys. Rev. B} {\bf 78} 195125.

\bibitem{roy08}
Roy R. 2008. Topological superfluids with time reversal symmetry. arXiv:0803.2868.

\bibitem{qiPRL09}
Qi XL, Hughes TL, Raghu S, Zhang SC. 2009. 
Time-Reversal-Invariant Topological Superconductors and Superfluids in Two and Three Dimensions.
{\it Phys. Rev. Lett.} {\bf 102} 187001.

\bibitem{volovikJETP09v2}
Volovik GE. 2009. Topological invariant for superfluid 3He-B and quantum phase transitions.
{\it JETP Lett.} {\bf 90} 587.

\bibitem{satoPRB09}
Sato M. 2009. Topological properties of spin-triplet superconductors and Fermi surface topology in the normal state.
{\it Phys. Rev. B} {\bf 79} 214526.

\bibitem{wangPRB11}
Wang Z, Qi XL, Zhang SC. 2011. 
Topological field theory and thermal responses of interacting topological superconductors.
{\it Phys. Rev. B} {\bf 84} 014527.

\bibitem{ryuPRB12}
Ryu S, Moore JE, Ludwig AWW. 2012. 
Electromagnetic and gravitational responses and anomalies in topological insulators and superconductors.
{\it Phys. Rev. B} {\bf 85} 045104. 

\bibitem{nomuraPRL12}
Nomura K, Ryu S, Furusaki A, Nagaosa N. 2012. 
Cross-Correlated Responses of Topological Superconductors and Superfluids.
{\it Phys. Rev. Lett.} {\bf 108} 026802.

\bibitem{satoPRB11}
Sato M, Tanaka Y, Yada K, Yokoyama T. 2011. 
Topology of Andreev bound states with flat dispersion.
{\it Phys. Rev. B} {\bf 83} 224511.

\bibitem{grover}
Grover T, Sheng DN, Vishwanath A. 2014.
Emergent Space-Time Supersymmetry at the Boundary of a Topological Phase.
{\it Science} {\bf 344} 280-283.

\bibitem{sasaki}
Sasaki S, Mizushima T. 2015.
Superconducting doped topological materials.
{\it Physica C} {\bf 514} 206-217.

\bibitem{asanoPRB14}
Asano Y, Fominov YV, Tanaka Y. 2014. 
Consequences of bulk odd-frequency superconducting states for the classification of Cooper pairs.
{\it Phys. Rev. B} {\bf 90} 094512.

\bibitem{higashitaniPRB12}
Higashitani S, Matsuo S, Nagato Y, Nagai K, Murakawa S, Nomura R, Okuda Y. 2012.
Odd-frequency Cooper pairs and zero-energy surface bound states in superfluid ${}^{3}$He.
{\it Phys. Rev. B} {\bf 85} 024524.


\bibitem{tsutsumi:2012c}
Tustsumi Y, Machida K. 2012.
Edge Current due to Majorana Fermions in Superfluid $^{3}$He A- and B-Phases.
{\it J. Phys. Soc. Jpn.} {\bf 81} 074607.


\bibitem{buchholtz:1981}
Buchholtz LJ, Zwicknagl G. 1981. 
Identification of $p$-wave superconductors.
{\it Phys. Rev. B} {\bf 23} 5788.

\bibitem{choi}
Choi H, Davis JP, Pollanen J, Halperin WP. 2006. 
Surface Specific Heat of $^{3}\mathrm{He}$ and Andreev Bound States.
{\it Phys. Rev. Lett.} {\bf 96} 125301.

\bibitem{bunkov}
Bunkov Y, Gazizulin R. 2015.
Majorana Fermions: Direct Observation in $^3$He
arXiv:1504.01711v1.

\bibitem{davisPRL08}
Davis JP, Pollanen J, Choi H, Sauls JA, Halperin WP, Vorontsov AB. 2008.
Anomalous Attenuation of Transverse Sound in $^{3}\mathrm{He}$.
{\it Phys. Rev. Lett.} {\bf 101} 085301.





\end{thebibliography}
\end{document}